\documentclass[useAMS,usenatbib]{mnras}
\usepackage{graphicx}
\usepackage{epsfig}
\usepackage{rotating}
\usepackage{amssymb}
\usepackage{color}
\usepackage{fancyhdr}
\usepackage{textcomp}
\usepackage{longtable}
\usepackage{tabularx}
\usepackage{url}

\def \lephare	{{\it Le PHARE \,}}

\title[LoCuSS: Weakly-lensed blue galaxies]{LoCuSS: Exploring the selection of faint blue background galaxies for cluster weak-lensing}
\author[F. Ziparo et al.]
       {Felicia Ziparo$^1$\thanks{E-mail: fziparo@star.sr.bham.ac.uk}, 
         Graham P. Smith$^1$, 
         Nobuhiro Okabe$^{2,3}$, 
         Chris P. Haines$^4$,\and
         Maria J. Pereira$^5$,
         Eiichi Egami$^5$\\ \\
         $^1$ School of Physics and Astronomy, University of Birmingham,
         Edgbaston, Birmingham B15 2TT, England \\
	 $^2$ Department of Physical Science, Hiroshima University, 1-3-1
	 Kagamiyama, Higashi-Hiroshima, Hiroshima 739-8526, Japan \\
	 $^3$ Kavli Institute for the Physics and Mathematics of the Universe (WPI), Todai Institutes for Advanced Study,University of Tokyo, 5-1-5 Kashiwanoha, Kashiwa, Chiba 277-8583, Japan \\
         $^4$ Departamento de Astronom\'ia, Universidad de Chile, Casilla 36-D, Correo Central, Santiago, Chile\\
         $^5$ Steward Observatory, University of Arizona, 933 North Cherry Avenue, Tucson, AZ 85721, USA
       }

\def\Om{\mathrel{\Omega_{\rm M}}}
\def\Ol{\mathrel{\Omega_\Lambda}}
\def\fwhm{\mathrel{\rm FWHM}}
\def\zphot{\mathrel{z_{\rm phot}}}
\def\zspec{\mathrel{z_{\rm spec}}}
\def\keV{\mathrel{\rm keV}}
\def\kpc{\mathrel{\rm kpc}}
\def\Mpc{\mathrel{\rm Mpc}}
\def\kms{\mathrel{\rm km\,s^{-1}}}
\def\degree{\mathrel{\rm degree}}

\def\pbkg{\mathrel{P_{\rm bkg}}}
\def\pbkgbin{\mathrel{P_{\rm bkg}^{\rm bin}}}
\def\zmax{\mathrel{z_{\rm max}}}
\def\ls{\mathrel{\hbox{\rlap{\hbox{\lower4pt\hbox{$\sim$}}}\hbox{$<$}}}}
\def\gs{\mathrel{\hbox{\rlap{\hbox{\lower4pt\hbox{$\sim$}}}\hbox{$>$}}}}

\begin{document}

\date{Accepted... Received... ; in original form...}

\pagerange{\pageref{firstpage}--\pageref{lastpage}} \pubyear{2015}

\maketitle

\label{firstpage}

\begin{abstract}
  Cosmological constraints from galaxy clusters rely on accurate
  measurements of the mass and internal structure of clusters. An
  important source of systematic uncertainty in cluster mass and
  structure measurements is the secure selection of background
  galaxies that are gravitationally lensed by clusters. This issue has
  been shown to be particular severe for faint blue galaxies.  We
  therefore explore the selection of faint blue background galaxies,
  by reference to photometric redshift catalogs derived from the
  COSMOS survey and our own observations of massive galaxy
  clusters at $z\sim0.2$.  We show that methods relying on photometric
  redshifts of galaxies in/behind clusters based on observations
  through five filters, and on deep 30-band COSMOS photometric
  redshifts are both inadequate to identify safely faint blue
  background galaxies. This is due to the small number of filters
    used by the former, and absence of massive galaxy clusters
  at redshifts of interest in the latter. We therefore develop a
  pragmatic method to combine both sets of photometric redshifts to
  select a population of blue galaxies based purely on photometric
  analysis.  This sample yields stacked weak-lensing results
  consistent with our previously published results based on red
  galaxies.  We also show that the stacked clustercentric number
  density profile of these faint blue galaxies is consistent with
  expectations from consideration of the lens magnification signal of
  the clusters. Indeed, the observed number density of blue background
  galaxies changes by $\sim10-30$ per cent across the radial range
  over which other surveys assume it to be flat. 
\end{abstract}



\section{Introduction}

Weak gravitational lensing is a direct probe of the projected total
mass distribution in galaxy clusters, and is therefore a promising
technique for measuring the massive end of the halo mass function
  \citep{Hoekstra2013}.  Such measurements play an important role in
a broad range of cosmological studies \citep{Allen2011}, which in
  turn drives an increasing focus on systematic biases in galaxy
  cluster mass measurements
  \citep[e.g.][]{Okabe2013,Applegate2014,Hoekstra2015,Okabe2015}.
Whilst lensing benefits from the advantage of not requiring
assumptions about the nature and physical state of the matter within
clusters, there are three principal sources of systematic error: those
relating to data reduction and faint galaxy shape measurement, those
relating to selection of background galaxies and characterising their
redshift distribution, and those relating to the modelling of the
shear profile and measuring the underlying halo mass.

Accurate galaxy shape measurements and modelling of cluster mass
distributions have both benefited recently from simulations.  Building
on the initiative taken by the Shear TEsting Programme
\citep[STEP][]{Heymans2006,Massey2007}, faint galaxy shapes can be
measured for cluster weak-lensing to an accuracy of $\ls10\%$.  Indeed
some of the more accurate methods are able to achieve few per cent
systematics for galaxies as faint as $i\simeq25$, and also extend the
parameter space explored by STEP to include that which is relevant to
clusters, i.e.\ reduced shear of $g\simeq0.2-0.3$
\citep[e.g.][]{Okabe2013}.  On the mass modelling side, studies based
on cosmological numerical simulations have shown that the ensemble
mass calibration of galaxy cluster samples can be recovered to sub-5\%
accuracy, paying careful attention to modelling details including the
range of cluster centric radii over which models are fitted
\citep{Bahe2012, Becker2011}.

Accurate selection of background galaxies is arguably trickier than
the other two sources of systematic error due to the requirement to
estimate a robust redshift for a very large number of galaxies many of
which are fainter than the limit of the deepest spectroscopic redshift
surveys, i.e.\ $\sim23-26$th magnitude.  Early studies selected faint
galaxies in a single photometric band, arguing that faint cluster
galaxies are a sub-dominant population \cite[e.g.][]{Kneib2003,
  Smith2005}.  More recently, colour-magnitude diagrams have
  been used to exclude galaxies that lie on or close to the ridge line
  of cluster galaxies -- the so-called red sequence
  \cite[e.g.][]{Okabe2010, Hoekstra2012, Applegate2014}, and
  colour-colour plots have also been used to separate cluster and
  background galaxy populations \cite[e.g.][]{Limousin2007,
    Medezinski2010, Umetsu2010, High2012, Israel2012}.  Taking a step
  further, some authors have attempted to use photometric redshifts of
  faint galaxies, based on upto 5 photometric bands, e.g.\ $ugriz$ or
  $BVRiz$, to select background galaxies
  \cite[e.g.][]{Limousin2007,Gavazzi09,Gruen14,Applegate2014,
    Covone2014,McCleary15,Melchior15}.  Given the uncertainties on
  redshift estimates based on photometry and the faintness of the
  galaxies in question, none of these methods delivers catalogues of
  faint galaxies that are free from contamination from foreground and
  cluster galaxies.  Moreover, it is challenging to estimate
  accurately the contamination level inherent in any given method.

Contamination is generally dominated by faint cluster members, with a
cluster centric radial number density profile that is expected to rise
towards the cluster centre.  A number density profile of background
galaxies that is a declining function of cluster centric radius is
therefore interpreted as a signature of contamination.
\cite{Kneib2003} were the first, to our knowledge, to invoke the
assumption that the intrinsic (i.e. uncontaminated) number density
profile of background galaxies is flat, and to apply a statistical
correction to the measured shear profile to correct for the effects of
contamination.  This approach has returned to vogue recently
  \citep{Applegate2014,Hoekstra2012,Hoekstra2015}.  If the assumption
  of a flat number density profile of background galaxies is valid for
  an individual cluster (or an ensemble of clusters when modeling all
  clusters simultaneously), and if the lens magnification has a
  negligible effect on the number density of background galaxies, then
  in principle it should be possible to correct statistically for
  contamination in this way.

Another approach is to devise colour selection criteria
based on photometric redshift catalogues of deep and/or wide blank
field surveys.  For example early studies used the Hubble Deep Field
observations for this purpose; latterly it is more common to use one
of the Cosmological Evolution Survey \cite[COSMOS, e.g.][]{Capak2007,
  Ilbert2006} or CFHT-LS \citep{Hildebrandt2012} photometric redshift
catalogues.  The main disadvantage of this approach is that, even if
the reference catalogue of redshifts is perfectly accurate,
these fields deliberately avoid lines of sight through massive galaxy
clusters.  Therefore any attempt to estimate the contamination
fraction using these catalogues is at best indicative due to the
absence, by design, of the troublesome contaminating galaxies at the
relevant cluster redshifts.  Nevertheless, these methods have been
important to improve the accuracy of background galaxy selection.

It is also possible to exploit the lensing signal itself to
characterise contamination.  These methods rest on the fact that to
first order the shapes of contaminating galaxies are uncorrelated with
the cluster shear signal and therefore simply reduce the measured
signal.  Specifically, the reduced tangential shear measured in a
given radial bin centred on a cluster is the error weighted sum of
reduced tangential shear of each galaxy divided by the sum of the
weights.  Contaminating galaxies therefore contribute no signal to the
numerator and increase the denominator through the weight that is
assigned to them.  Typically the weight function reflects the
measurement uncertainty on the galaxy shape, and not the probability
that the galaxy is indeed behind the cluster.  The maximal lensing
signal is therefore measured from the least contaminated sample of
galaxies.  \cite{Medezinski2007} therefore introduced the idea of
measuring the shear signal of a cluster as a function of colour cut
used to define the background galaxy sample, aiming to identify a
colour cut beyond which the shear signal saturates.  \cite{Okabe2010}
applied these methods to 30 galaxy clusters, selecting galaxies redder
and bluer than the cluster sequence, for use as background galaxies.
Subsequently \citet{Okabe2013} combined this approach with an
  analytic model of the contamination fraction to achieve 1 per cent
  contamination in samples of galaxies redder than the cluster red
  sequence by a colour offset of $\Delta(V-i)\ge0.475$.  Whilst this
  approach delivered very pure background galaxy samples, it yielded
  just $5$ background galaxies per square arcminute per cluster,
  i.e.\ insufficient to measure the mass of individual clusters in
  their sample.  \citeauthor{Okabe2013} also found that their
  techniques produced ambiguous results for blue galaxies, largely
  because the relationship between colour and redshift is more
  complicated for blue galaxies than for red galaxies.  In a companion
  to this article, \cite{Okabe2015} extend \citeauthor{Okabe2013}'s
  red galaxy selection methods to incorporate a colour-cut that
  depends on clustercentric radius, and thus achieve a number density
  of 13 galaxies per square arcminute per cluster, again with just 1
  per cent contamination.

In this article we investigate several of the issues highlighted above
in the context of the Local Cluster Substructure Survey
(LoCuSS\footnote{\url{http://www.sr.bham.ac.uk/locuss}}).  Our main
objective is to explore the selection of blue background galaxies in
the $(V-i)/i$ colour-magnitude plane.  As outlined above, it is now
clear that methods that rely on ``saturation'' of the shear signal are
unreliable for blue galaxies that are selected in this plane.  We have
therefore obtained $BVRiz$-band data for a sub-sample of clusters to
study the relationship between photometric redshift, $(V-i)$ colour,
and apparent $i$-band magnitude.  The main outcome for our own
programme is therefore to assess the reliability of blue galaxy
selection.  We judge our results based on the target to
  control systematic biases in LoCuSS weak-lensing mass measurements
  at the sub-4 per cent level, as set out in \cite{Okabe2015}.  We
  therefore ask the specific question: can we control the
  contamination of blue background galaxy catalogues at this level?
  Our approach is based on careful testing of the accuracy of
  photometric redshifts, and detailed investigation of the limitations
  of photometric redshifts based on just five photometric bands and/or
  on blank field observations.  This article therefore has broad and
  significant relevance for the community, in addition to helping us
  to achieve the goals of LoCuSS.

In \S\ref{sec:data} we describe the data and the clusters we use for
this analysis; in \S\ref{sec:analysis} we present the photometric
redshift calculation that we then use to define our background galaxy
selection method in \S\ref{sec:results}.  We discuss our results and
conclusions in \S\ref{sec:discussion}.  Throughout our analysis we
adopt the following cosmological parameters: $H_0=70\kms\Mpc^{-1}$,
$\Ol=0.7$, $\Om=0.3$.  All magnitudes are in the AB system.


\section[]{Data}\label{sec:data}

We use a sample of 50 clusters from the Local Cluster Substructure
Survey (LoCuSS), for which sensitive high angular resolution imaging
data are available from Suprime-Cam \citep{Miyazaki2002} on the Subaru
8.2-m telescope\footnote{Based in part on observations obtained at the
  Subaru Observatory under the Time Exchange program operated between
  the Gemini Observatory and the Subaru Observatory.}\footnote{Based
  in part on data collected at Subaru Telescope and obtained from the
  SMOKA, which is operated by the Astronomy Data Center, National
  Astronomical Observatory of Japan.}.  The clusters comprise the
so-called ``High-$L_X$'' sample and satisfy the following selection
from the \emph{ROSAT} All Sky Survey catalogues
\citep{Ebeling1998,Ebeling2000,Boehringer2004} $0.15\le z\le0.30$,
$n_H\le7\times10^{-20}{\rm cm}^{-2}$,
$-25^\circ\le\delta\le+65^\circ$,
$L_{X[0.1-2.4\keV]}/E(z)\ge4.1\times10^{44}{\rm erg\,s}^{-1}$, where
$E(z)=(\Om(1+z)^3+\Ol)^{0.5}$.  This is the same sample as
  discussed by \citet{Okabe2013}, \citet{Martino2014}, and
  \citet{Okabe2015}.

Full details of the Subaru observations, data reduction, photometric
calibration, and faint galaxy shape measurements are given by
\citet{Okabe2013} and \cite{Okabe2015}.  In summary,
forty seven clusters were observed through the $V$- and $i'$-band
filters, and the remaining three through the $V/I_{\rm C}$, $g/i'$,
and $B/i'$-band filters.  Our results are insensitive to whether we
include or exclude these three clusters.  Hereafter we refer to the
redder filter as $i$-band and the bluer filter as $V$-band.  The
$i$-band data were obtained in excellent conditions, with $\fwhm_{\rm
  median}=0.7''$, and typically reach a $5\sigma$ point source
sensitivity of $i_{\rm AB}=26$.

We have also observed three of the fifty clusters through $B$, $R$,
and $z$-band filters to allow us to measure photometric redshifts of
faint galaxies for this study.  These clusters are ABELL0068,
ABELL0383 and ABELL0611; they lie at redshifts $z=0.251$, $z=0.188$
and $z=0.288$, respectively.  We have also observed these three
clusters with Hectospec \citep{Fabricant2005} mounted on the Multiple
Mirror Telescope (MMT) at Mount Hopkins, Arizona, as part of the
Arizona Cluster Redshift Survey (ACReS; Pereira et al., in
preparation; see also \citealt{Haines2013}).  ABELL0383 is also
covered by the redshift survey of \cite{Geller2014}.  We assemble a
total of 2390 secure spectroscopic redshifts of which 796 are
confirmed cluster members.  We use these redshifts to train our
photometric redshift measurements in \S\ref{sec:photoz}.


\section[]{Analysis}
\label{sec:analysis}

\subsection{Photometry}

We use the $V/i$-band photometric catalogues for all 50 clusters from
\citet{Okabe2013} and described in detail by \cite{Okabe2015}.
Matched $BRz$-band catalogues are constructed following the same
procedures.  We adopt the SExtractor \citep{Bertin1996} parameter {\sc
  mag\_auto} as the total magnitude of each object in the $i$-band,
and measure aperture magnitudes in all bands within apertures matched
to $1.5\times$ the worst seeing ($\sim 0.85\arcsec$) of the available
filters for each cluster, after smoothing each frame to the worst
seeing.  This approach is well matched to the photometric analysis of
COSMOS observations \citep{Ilbert2006}.

We check the reliability of the photometric calibration in two ways.
First, we select unsaturated stars with $i>20$ and {\sc
    class\_star}$\ge0.99$, and compare their colours with those
derived from template spectra from \cite{Pickles1998}, white dwarves
and other stellar types.  The observed and synthetic colours
  match very well and imply residual uncertainties on the photometric
  calibration at the sub-10 per cent level -- i.e.\ consistent with
  the calibration analysis of \cite{Okabe2015}.  Second, we perform a
more quantitative test, using \lephare (PHotometric Analysis for
Redshift Estimations; \citealt{Arnouts2001, Ilbert2006}) in ``training
mode''.  For objects of known redshift \lephare recursively computes
the systematic offsets for the photometric zero-points \cite[see
][]{Ilbert2006}.  The resulting shifts are of the order of a few per
cent.


\subsection{Photometric redshifts}
\label{sec:photoz}

We compute photometric redshifts for all sources detected at $i\le26$
in the Suprime-Cam observations of ABELL0068, ABELL0383 and ABELL0611,
using ten Spectral Energy Distribution (SED) templates based on
multi-wavelength observations of Virgo cluster \citep{Boselli2003} and
twelve starburst galaxy templates used by COSMOS \citep{Ilbert2009},
and generated with \cite{BC03} stellar population synthesis models.
The Virgo templates best represent cluster galaxies at $z\simeq0.2$
\citep{Ziparo2012}; the COSMOS templates span the range of spectral
properties expected of higher redshift galaxies.  Dust extinction is
applied to all templates bluer than an Sa spectral type using the
modified \cite{Calzetti2000} attenuation law with $0\le E(B-V)\le0.5$
and with a step of $\Delta E(B-V)=0.1$. Finally, we add a set of emission
lines to the templates following \cite{Ilbert2006}.

\begin{figure*}
\centering
  \includegraphics[trim=3cm 5cm 1cm 4cm,clip=true,angle=-90, width=0.48\hsize]{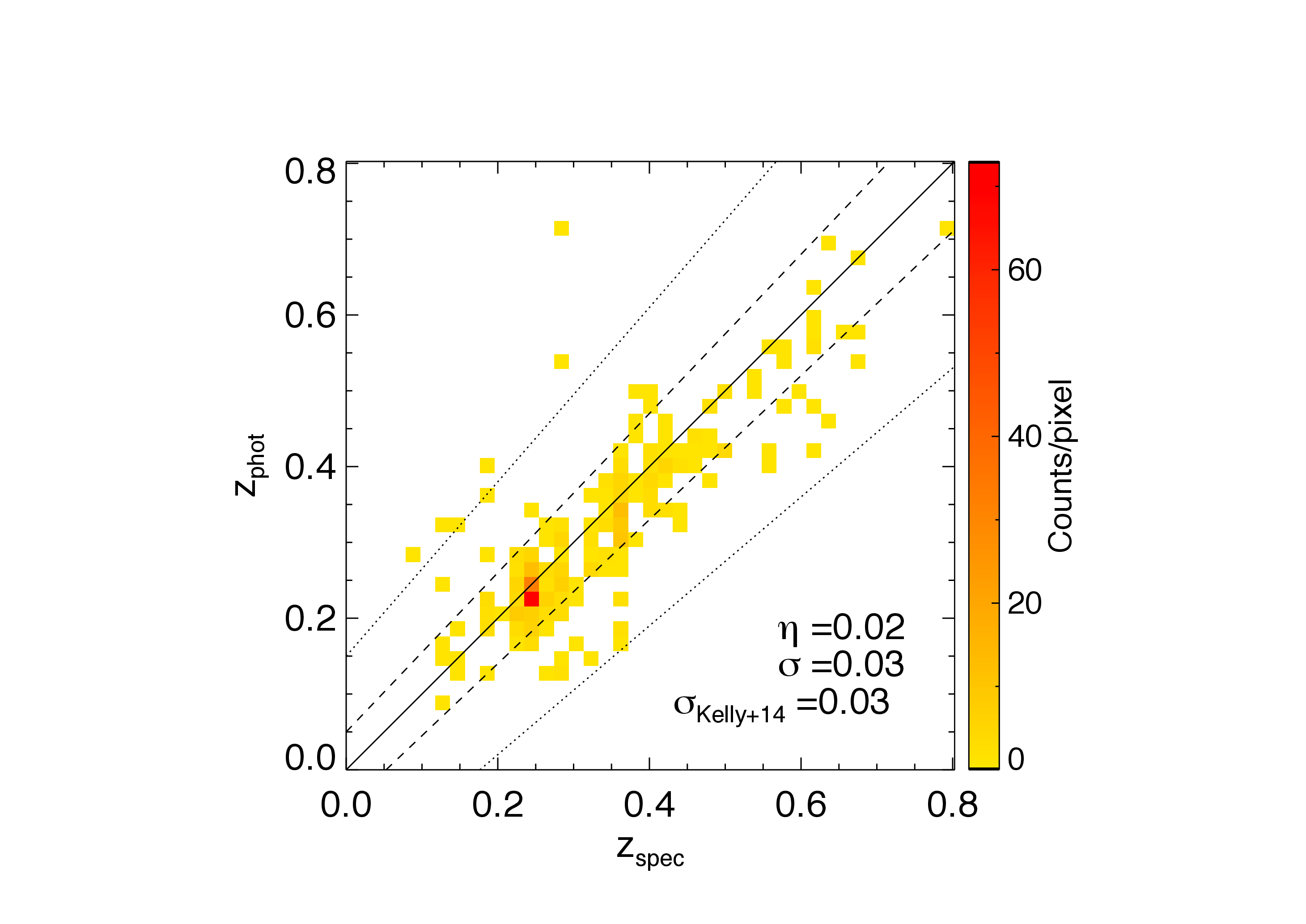}
  \includegraphics[trim=5.2cm 7cm 4cm 8cm,clip=true,angle=-90, width=0.48\hsize]{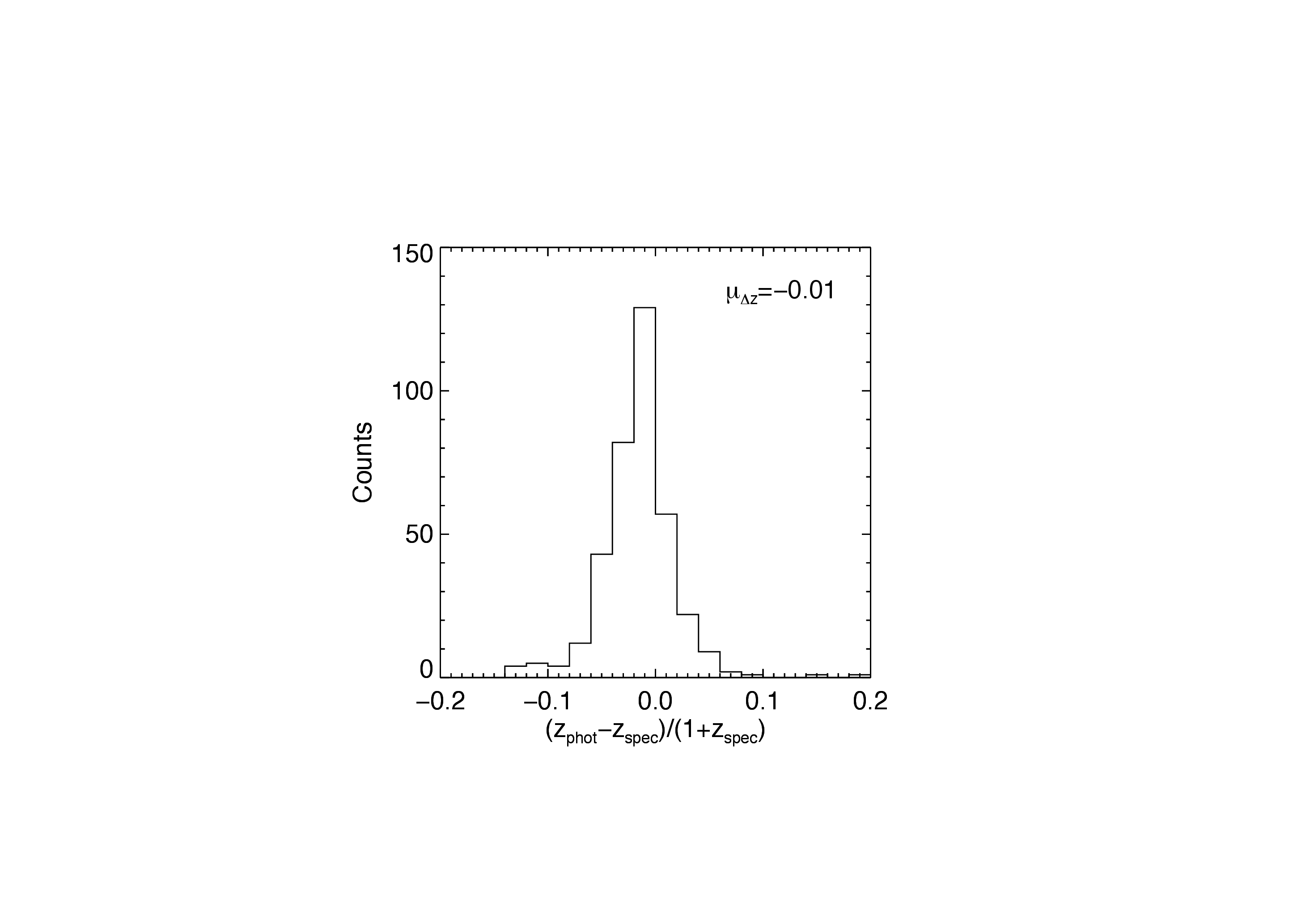}
  \includegraphics[trim=3cm 5cm 1cm 4cm,clip=true,angle=-90, width=0.48\hsize]{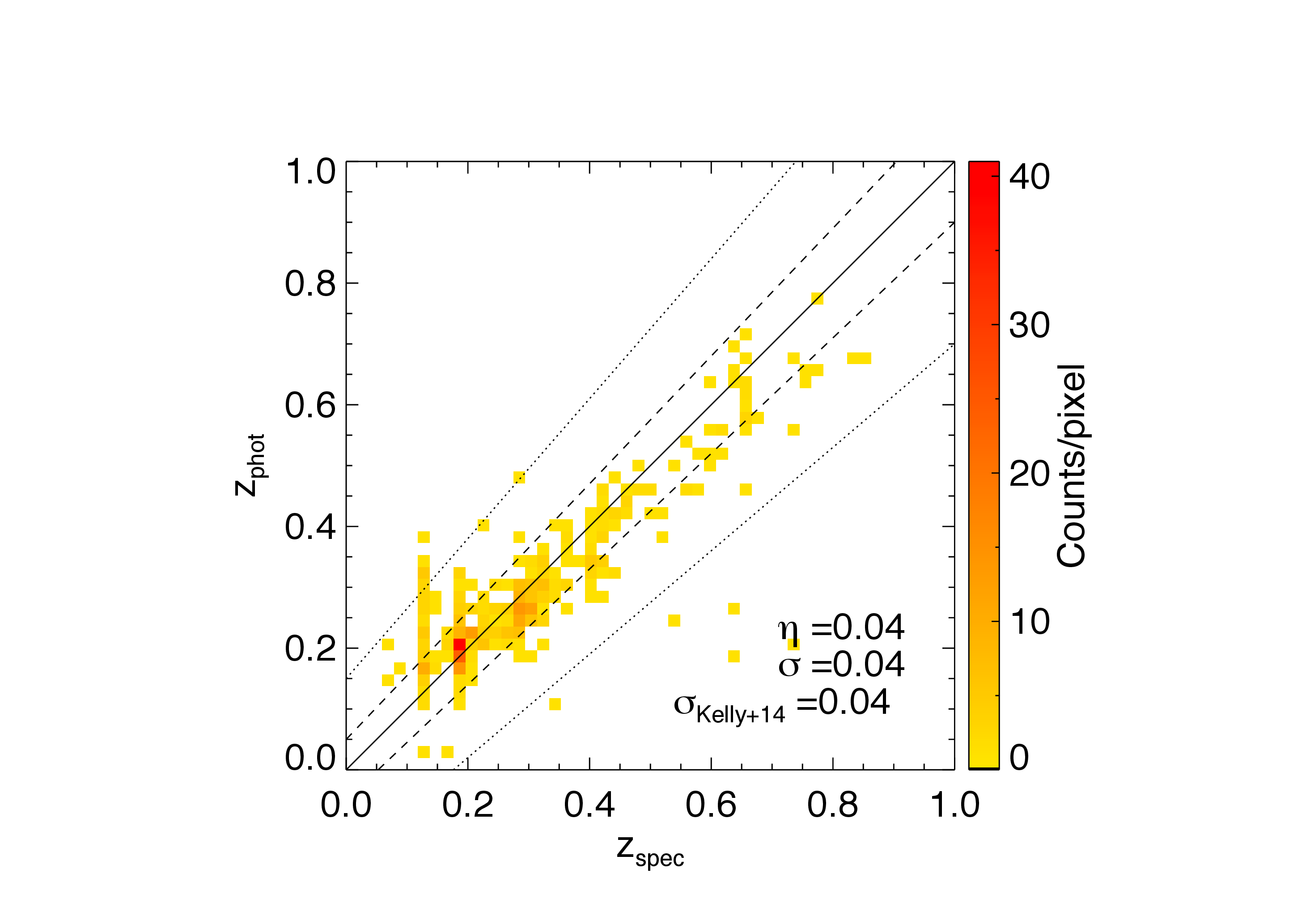}  
  \includegraphics[trim=5.2cm 7cm 4cm 8cm,clip=true,angle=-90, width=0.48\hsize]{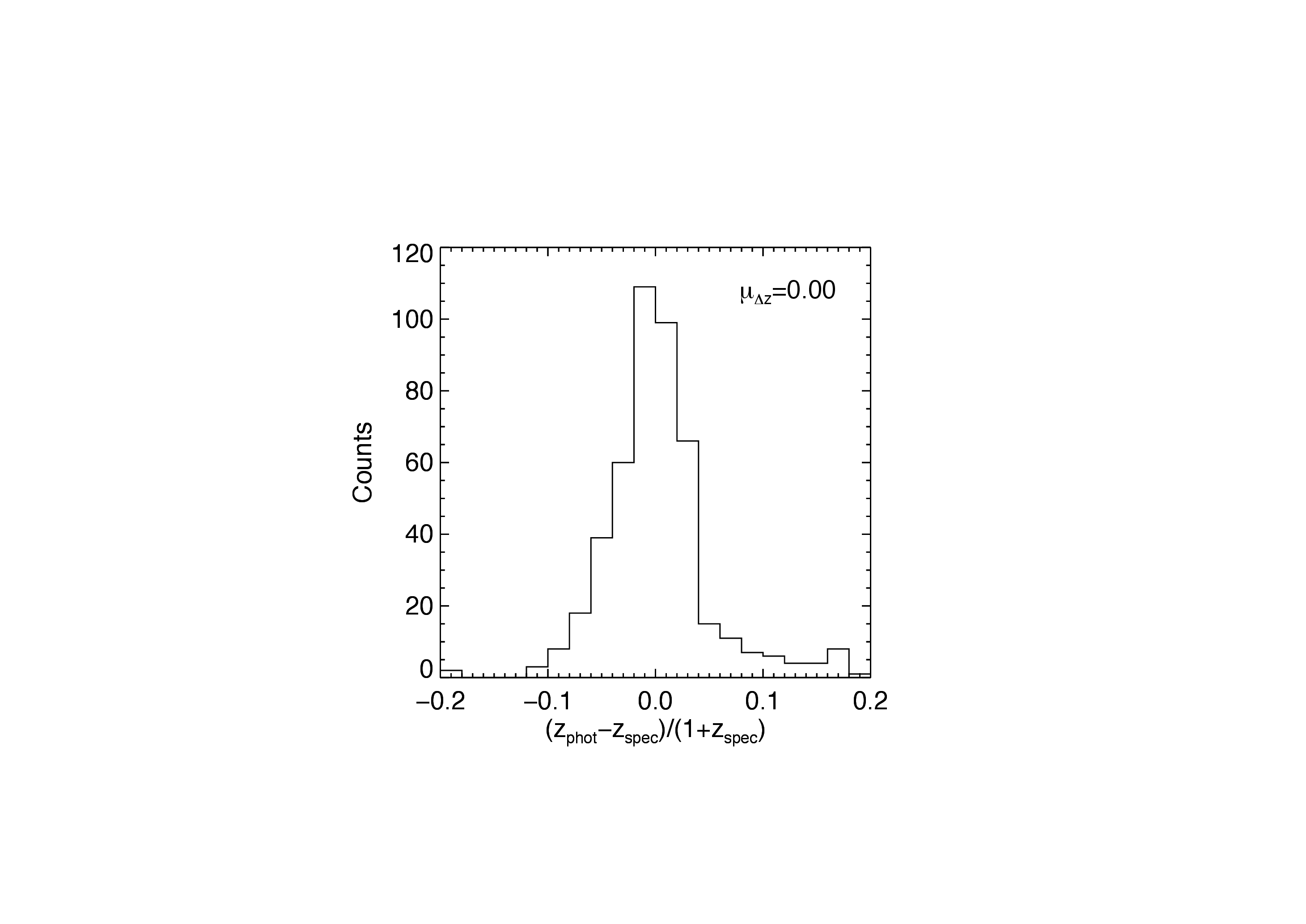}  
  \includegraphics[trim=3cm 5cm 1cm 4cm,clip=true,angle=-90, width=0.48\hsize]{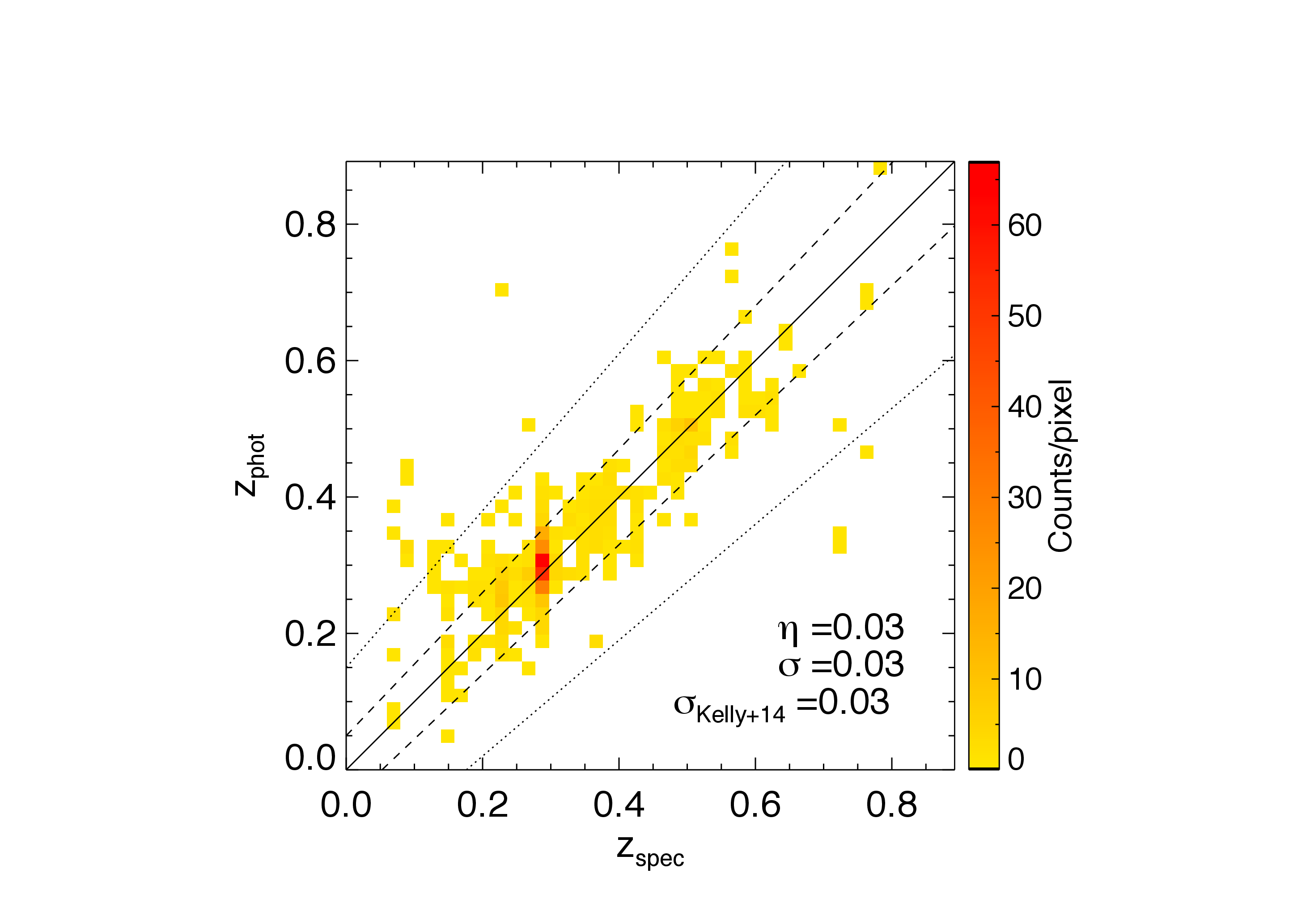}  
  \includegraphics[trim=5.2cm 7cm 4cm 8cm,clip=true,angle=-90, width=0.48\hsize]{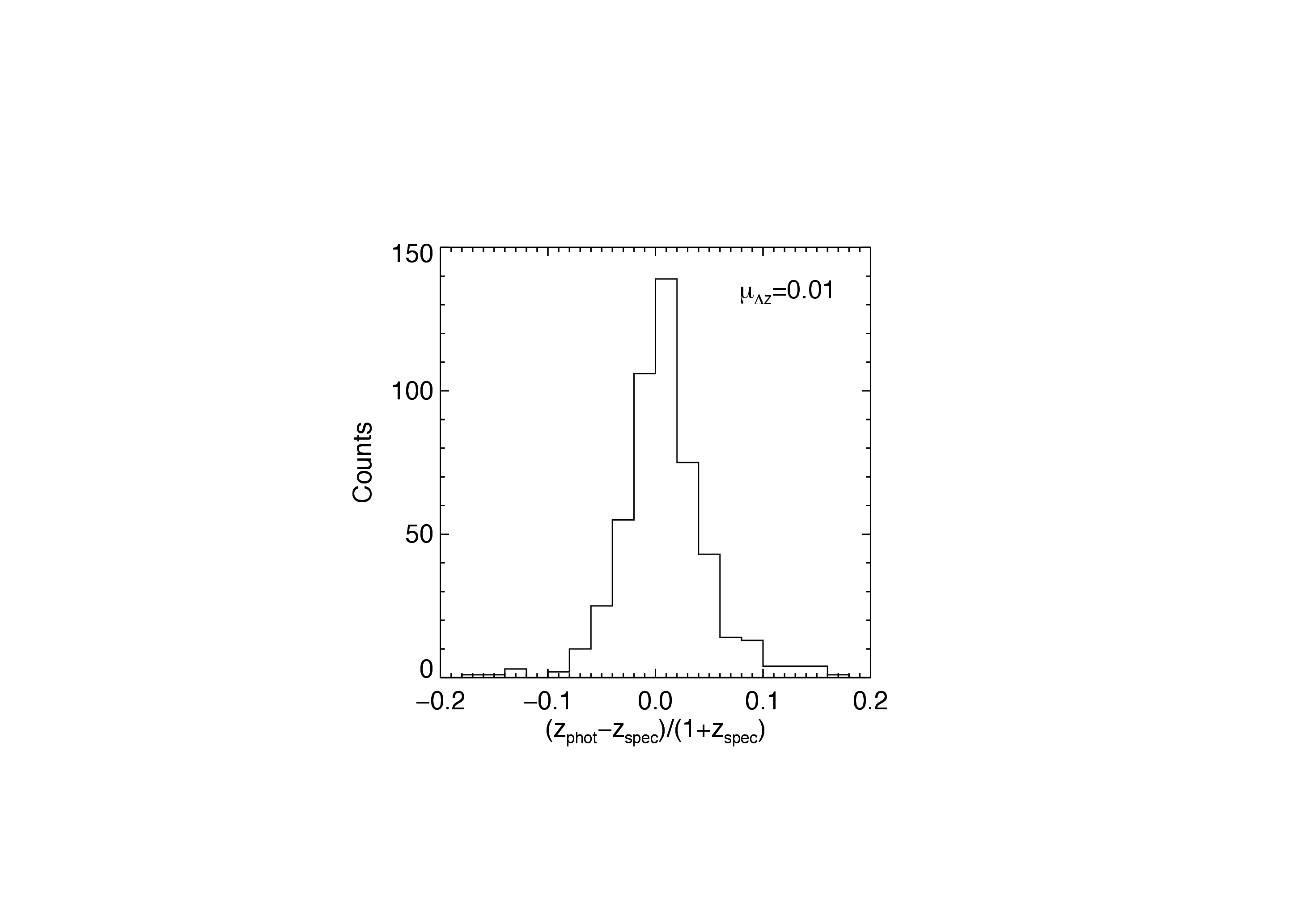}  
  \caption{Photometric ($\rm z_{phot}$) versus spectroscopic redshift
    ($\rm z_{phot}$, left panels) and distribution of $\rm
    (z_{phot}-z_{spec})/(1+z_{spec})$ (right panels) for ABELL0068,
    ABELL0383 and ABELL0611 in the top, middle and bottom panels,
    respectively.  The solid lines in the left panels represent $\rm
    z_{spec}=z_{phot}$, while dashed and dotted lines are for
    $\mathrm{z_{phot}=z_{spec}\pm0.05(1+z_{spec})}$ and
    $\mathrm{z_{phot}=z_{spec}\pm0.15(1+z_{spec})}$,
    respectively. $\rm \eta$ is the fraction of catastrophic failures,
    $\rm \sigma$ represents the accuracy as computed by
    \citet{Ilbert2006} and $\rm \sigma_{Kelly+14}$ is the accuracy as
    computed by \citet{Kelly2012}. The colour bar in the left panels
    shows the density of points calculated in counts/pixels with red colour highlighting that galaxies are
        concentrated at the cluster redshift. $\rm \mu_{\Delta z}$
    is the mean of the distribution in the right panels. }
  \label{fig:cluster_photoz}
\end{figure*}

We test the reliability of the photometric redshifts by comparing them
with spectroscopic redshifts from ACReS (\S\ref{sec:data}), defining
the catastrophic failure rate, $\eta$, following
\citeauthor{Ilbert2006}, as the fraction of objects for which
$|{\zphot}-{\zspec}|/(1+{\zspec})>0.15$.  We also compute the accuracy
as:
\begin{equation}
\sigma=1.48\times{\rm median}(|{\zphot}-{\zspec}|/(1+{\zspec}))
\end{equation}
following \citet{Ilbert2006}.  We note \citet{Kelly2012} adopt a
different definition of accuracy, preferring to exclude outliers from
their calculation.  To facilitate comparison with their work, we also
compute $\sigma_{\rm Kelly+14}$, which they define as: the standard
deviation of $\rm |({\zphot}-{\zspec})/(1+{\zspec})|$ after rejecting
the outliers with $|({\zphot}-{\zspec})/(1+{\zspec})|>0.1$.

We obtain a photometric redshift accuracy of $\sigma\simeq0.03-0.04$
and a catastrophic failure fraction of $\eta\simeq0.02-0.04$ (left
panels in Fig.~\ref{fig:cluster_photoz}); $\sigma_{\rm Kelly+14}$ is
consistent with $\sigma$, for this comparison with spectroscopic
redshifts.  We also compute $\mu_{\Delta z}$ to test for any
systematic over- or under-estimation of redshift, defined thus:
\begin{equation}
\mu_{\Delta z}=\left\langle\frac{{\zphot}-{\zspec}}{1+{\zspec}}\right\rangle
\end{equation}
constraining any systematic bias at the $\ls1\%$ level (right panels
of Fig.~\ref{fig:cluster_photoz}).  Overall, our tests suggest that
for galaxies with sufficiently high S/N, BVRiz photometric redshifts
are able to robustly distinguish contaminant cluster galaxies from
background ($z>0.4$) galaxies.


\subsection{Photometric redshifts beyond 20th magnitude}
\label{sec:photoz_COSMOS}

The median i-band magnitude of galaxies with ACReS redshift is
$i=19.38$, and 95\% of the sources have $i<20.8$.  Therefore, whilst
encouraging, the tests described in \S\ref{sec:photoz} do not
examine directly the reliability of galaxies used for
weak-lensing mass measurements.  We therefore turn to the COSMOS
UltraVISTA photometric redshift catalogue
\citep{McCracken2012,Ilbert2013} that is limited at $i<27.5$, and
benefits from four deep near-infrared filters $Y$, $J$, $H$ and $K_S$.
This filter coverage enables more robust photometric redshifts for
galaxies at $z>1.3$ than earlier versions of the COSMOS catalogue,
since the Balmer break is redshifted to the near-infrared for these
galaxies.  Furthermore, the COSMOS UltraVISTA photometric redshifts
are tested against almost 35,000 new spectra with galaxies at $z>1.5$
\citep[for more details see][]{Ilbert2013}.

\begin{figure*}
\centering
   \includegraphics[trim=3cm 5cm 1cm 4cm,clip=true,angle=-90,width=0.48\hsize]{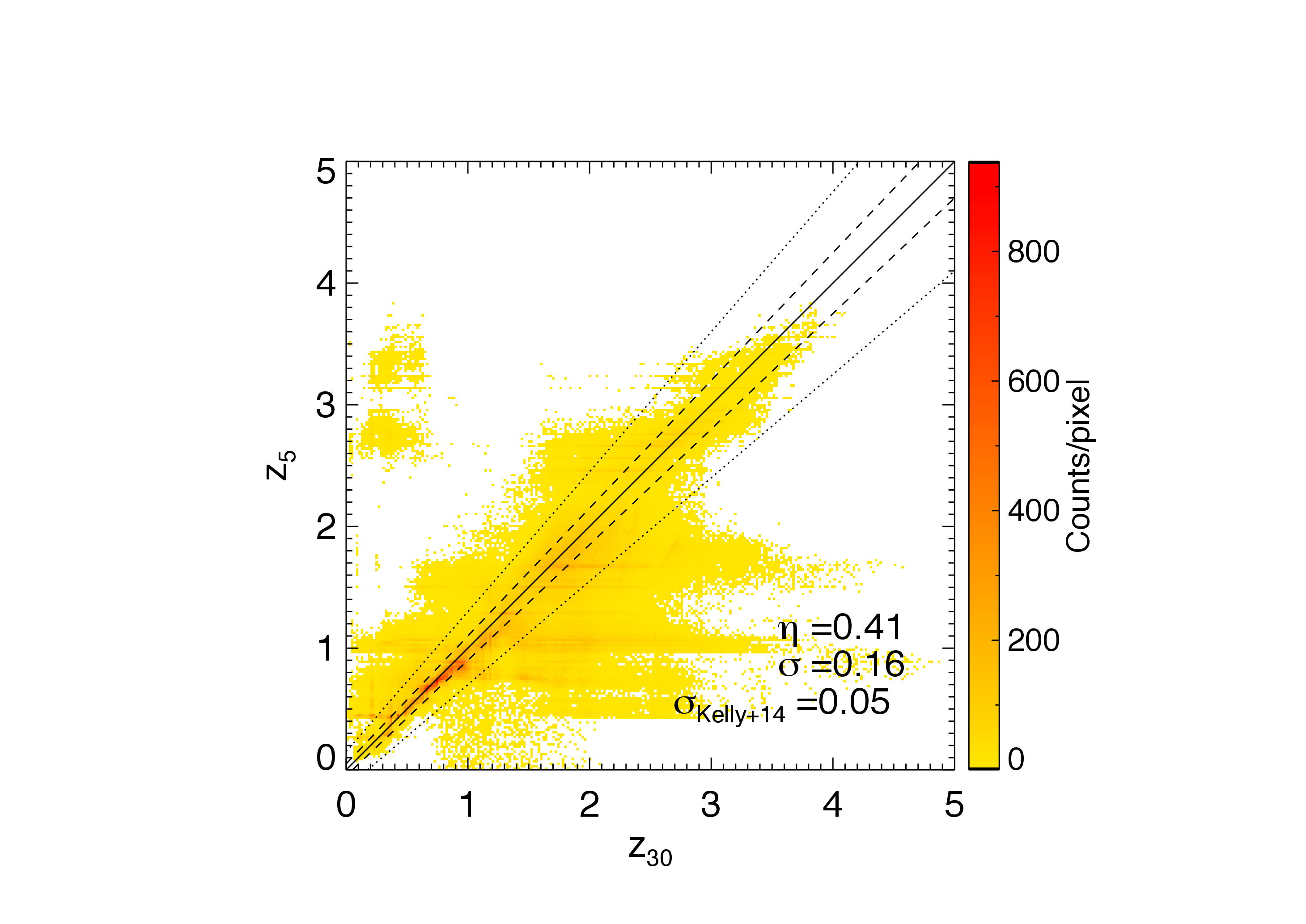}
   \includegraphics[trim=5.2cm 7cm 4cm 8cm,clip=true, angle=-90, width=0.48\hsize]{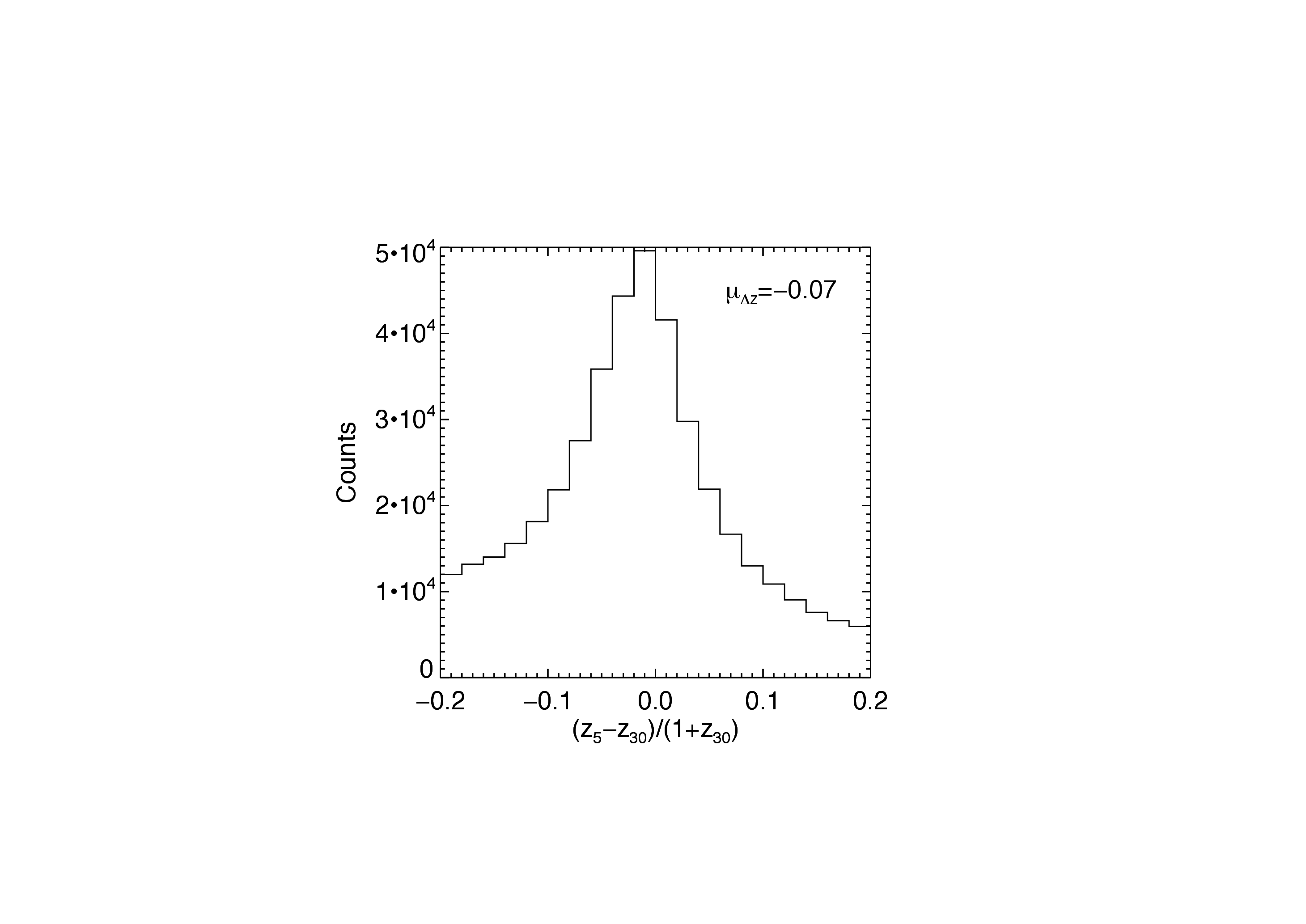}
   \includegraphics[trim=3cm 5cm 1cm 4cm,clip=true, angle=-90,width=0.48\hsize]{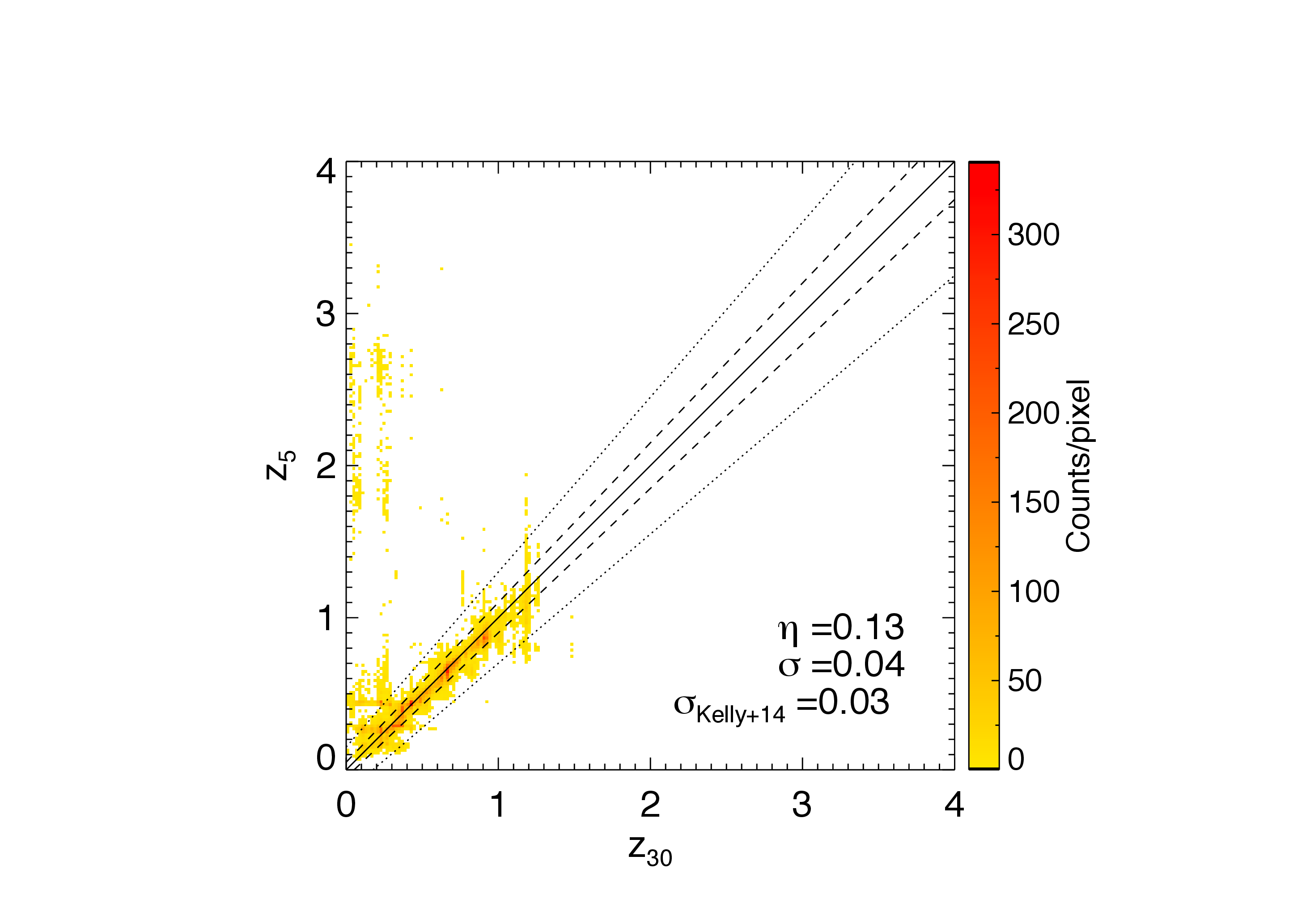}
   \includegraphics[trim=5.2cm 7cm 4cm 8cm,clip=true,angle=-90, width=0.48\hsize]{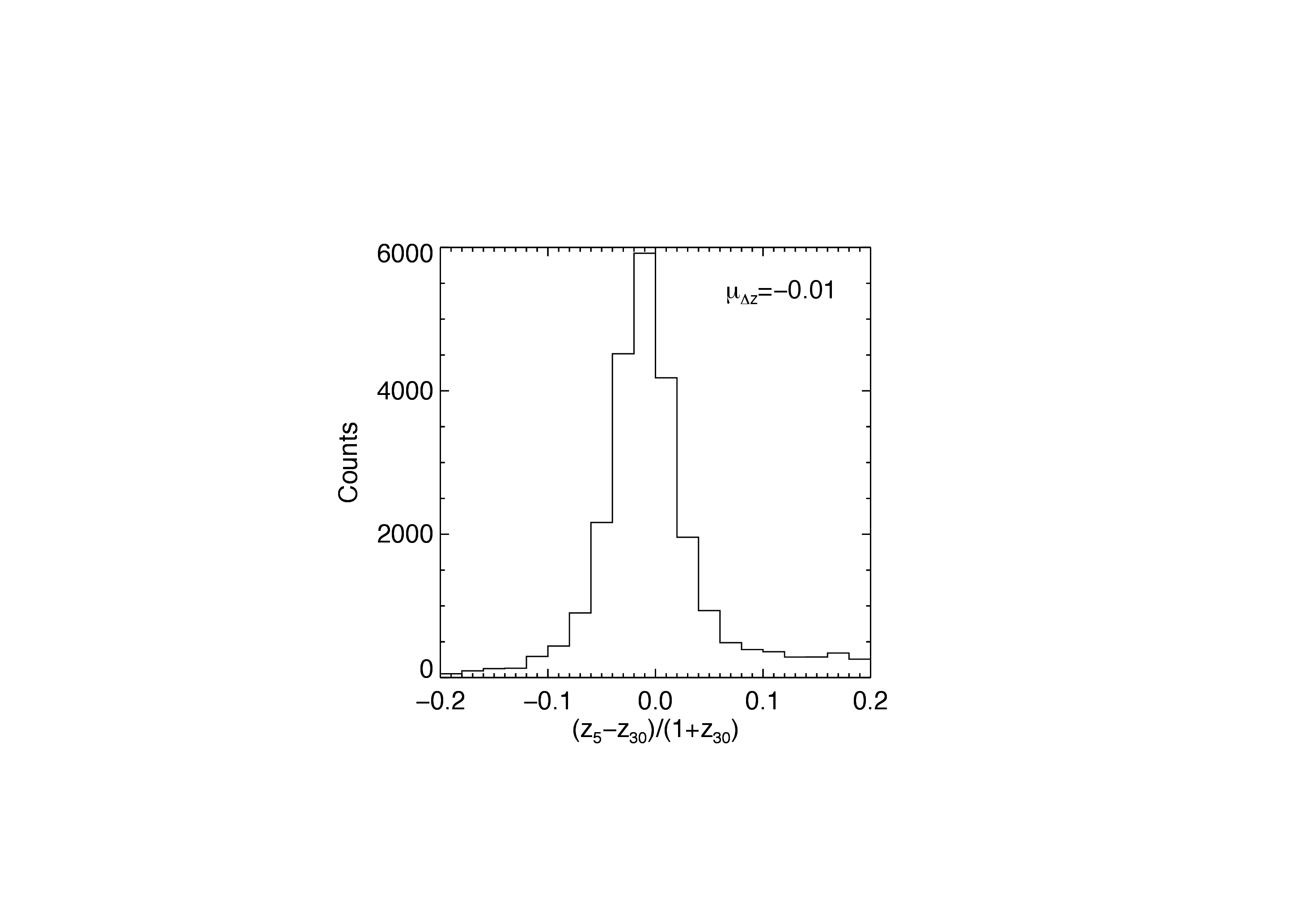}
   \caption{5-band photometric redshifts $z_5$ computed in this work
     versus 30-band photometric redshifts $z_{30}$ from
     \citet{Ilbert2013} for all galaxies (top left panel) and no more
     than one peak in the P(z) above a probability of 5\% (bottom lef
     panel) in the COSMOS field. The right panels represent the
     distribution of $\rm (z_{5}-z_{30})/(1+z_{30})$ from the left
     corresponding panels. All symbols are the same as in
     Fig.~\ref{fig:cluster_photoz}.}
  \label{fig:zphot_comp_COSMOS}
\end{figure*}

A further advantage of the COSMOS data is that their optical data were
obtained with Suprime-Cam at Subaru, allowing a straightforward
comparison with our analysis.  We therefore use the COSMOS photometry
and photometric redshifts (based on 30 photometric bands) to test our
method for computing photometric redshifts from $BVRiz$ photometry.
We apply the methods described in \S\ref{sec:photoz} to COSMOS $BVRiz$
photometry and compare the resulting COSMOS 5-band photometric
redshifts, $z_5$, with COSMOS 30-band photometric redshifts, $z_{30}$,
from \cite{Ilbert2013}.  Replacing $\zspec$ with $z_{30}$ in our
definitions of $\sigma$, $\eta$, and $\mu_{\Delta z}$, we obtain an
accuracy, catastrophic failure fraction, and systematic offset of:
$\sigma=0.16$, $\eta=0.41$, and $\mu_{\Delta z}=-0.07$ respectively
(Fig.~\ref{fig:zphot_comp_COSMOS}).  As expected, $BVRiz$ photometry
does not constrain photometric redshifts as well as 30 bands.  For
example, the cloud of points at $z_{30} <1$ and $z_{5}>2.5$ arise from
degeneracy between the 4000\,\AA{} and Lyman breaks, and
similarly at $z_{30}>1.5$.  We note that \cite{Kelly2012} performed a
similar test to ours, obtaining $\sigma_{\rm Kelly+14}=0.06$ down to
$i\simeq24$.  We obtain a similar result -- $\sigma_{\rm
  Kelly+14}=0.05$ -- if we follow their approach of excluding outliers
from the calculation of $\sigma$ (Fig.~\ref{fig:zphot_comp_COSMOS}).

An important caveat on the preceding analysis is that it compares one
set of photometric redshifts with another, not withstanding the fact
that the COSMOS filter set is the most complete in the literature at
these depths and solid angle.  We therefore perform a more refined
test of our 5-band photometric redshift methods, taking advantage of
the full redshift probability distribution, $P(z)$, from the COSMOS
catalogue.  \cite{Ilbert2006} noted that the catastrophic failure
fraction of galaxies that present a second $P(z)$ peak with a
probability greater than 5\% is $\eta=0.44$.  This is comparable with
our catastrophic failure fraction of $\eta=0.41$.  We therefore select
a ``pseudo-spectroscopic'' sub-sample of COSMOS galaxies that each
have a single $P(z)$ peak above a probability of 5\%, and obtain
$\sigma=0.04$, $\eta=0.13$, and $\mu_{\Delta z}=-0.01$ (bottom panels
of Fig.~\ref{fig:zphot_comp_COSMOS}).  This is similar to the
comparison with spectroscopic redshifts in \S\ref{sec:photoz}, albeit
with a higher catastrophic failure fraction.  Note that this
restricted sample of galaxies with single peaked $P(z)$ represents
just 4\% of the total COSMOS sample.

The inability of $BVRiz$-band photometric redshifts to correctly
locate some galaxies at $z\ls0.3$ that was noted above, is seen again
in the lower left panel of Fig.~\ref{fig:zphot_comp_COSMOS} -- note
the galaxies at $z_{30}<0.3$ and $z_{5}>1$.  However the effect is
less pronounced than when comparing $BVRiz$-band photometric redshifts
with the full COSMOS sample (upper left panel of
Fig.~\ref{fig:zphot_comp_COSMOS}), as is reflected in the improved
values of $\sigma$, $\eta$, and $\mu_{\Delta z}$.  Indeed, these
galaxies are generally blue, with $(V-i)\ls0.5$, and are typical of
the galaxies used in many galaxy cluster weak-lensing studies that use
colour and/or 5-band photometric redshift selection
\citep[e.g.][]{Okabe2010,Umetsu2010,High2012,Applegate2014,Hoekstra2015}.

\subsection{Summary}

We have computed photometric redshifts of galaxies detected within
the Suprime-Cam field of view, centred on three massive galaxy
clusters at $z\simeq0.2$.  These photometric redshifts are based on
deep $BVRiz$-band observations.  We use spectroscopic redshifts from
ACReS (\S\ref{sec:photoz}), and photometric redshifts from the COSMOS
UltraVISTA catalogue (\S\ref{sec:photoz_COSMOS}) to test the
reliability of these 5-band photometric redshifts, obtaining
encouraging results.  The accuracy is typically $\sigma\simeq0.04$,
the catastrophic failure fraction is $\eta\ls0.1$, and any systematic
bias is of the order $\mu_{\Delta z}\simeq0.01$.  Note that our tests
using the COSMOS catalogue use a ``pseudo-spectroscopic'' sub-sample
of COSMOS galaxies that have a single peak in their $P(z)$
distribution.  Nevertheless, the degeneracy between the spectral
shapes of blue galaxies at $z\ls0.3$ and $z\simeq2-3$ is not broken by
$BVRiz$-band photometry, causing concerns regarding contamination of
blue background galaxy samples selected based on these bands and
photometric redshifts derived from them.

\section{Results}
\label{sec:results}


We use the full $P(z)$ distribution of each galaxy in the three
photometric redshift catalogues discussed in \S\ref{sec:analysis} to
investigate the selection of background galaxies for weak-lensing
calculations.  These catalogues comprise: (1) ``LoCuSS'' $BVRiz$-band
photometric redshifts of galaxies along the line of sight through
ABELL0068, ABELL0383, and ABELL0611 (\S\ref{sec:photoz}); (2)
``COSMOS-30'' photometric redshifts of galaxies within the
$1.5\times1.5\degree^2$ footprint of the UltraVISTA observations of
the COSMOS field \citep{Ilbert2013}; (3) ``COSMOS-5'' $BVRiz$-band
photometric redshifts of galaxies in the same UltraVISTA footprint
(\S\ref{sec:photoz_COSMOS}).

We define $\pbkg$ as the probability that a galaxy is at a redshift of
$z\ge0.4$, i.e.\ well beyond the redshift of the clusters considered
here, and also beyond the redshift limit of the LoCuSS cluster sample
(\S\ref{sec:data}).  We compute $\pbkg$ for every galaxy in all three
catalogues as follows:
\begin{equation}
  \pbkg=\frac{\int_{0.4}^{\zmax}P(z)dz}{\int_0^{\zmax}P(z)dz}
  \label{p_z04}
\end{equation}
where $\zmax=6$ is the maximum redshift considered for the photometric
redshift calculations (\S\S\ref{sec:photoz}~\&~\ref{sec:photoz_COSMOS}).
In the following sections we investigate the utility of different
colour and magnitude cuts to define a sample of galaxies that
suffers minimal contamination by faint cluster and foreground
galaxies.  We therefore further define $\pbkgbin$ as the mean value of
$\pbkg$ for a given ``bin'' of $(V-i)/i$ colour-magnitude space:
$\pbkgbin=\langle\pbkg\rangle$.  This quantity allows us to link the
$P(z)$ information in the three photometric redshift catalogues with
the colour-magnitude information that is available for the full LoCuSS
sample.    


\subsection{Redshift as a function of colour and magnitude}
  
\begin{figure*}
  \centerline{
    \hspace{-5mm}
    \includegraphics[angle=-90,width=0.95\hsize]{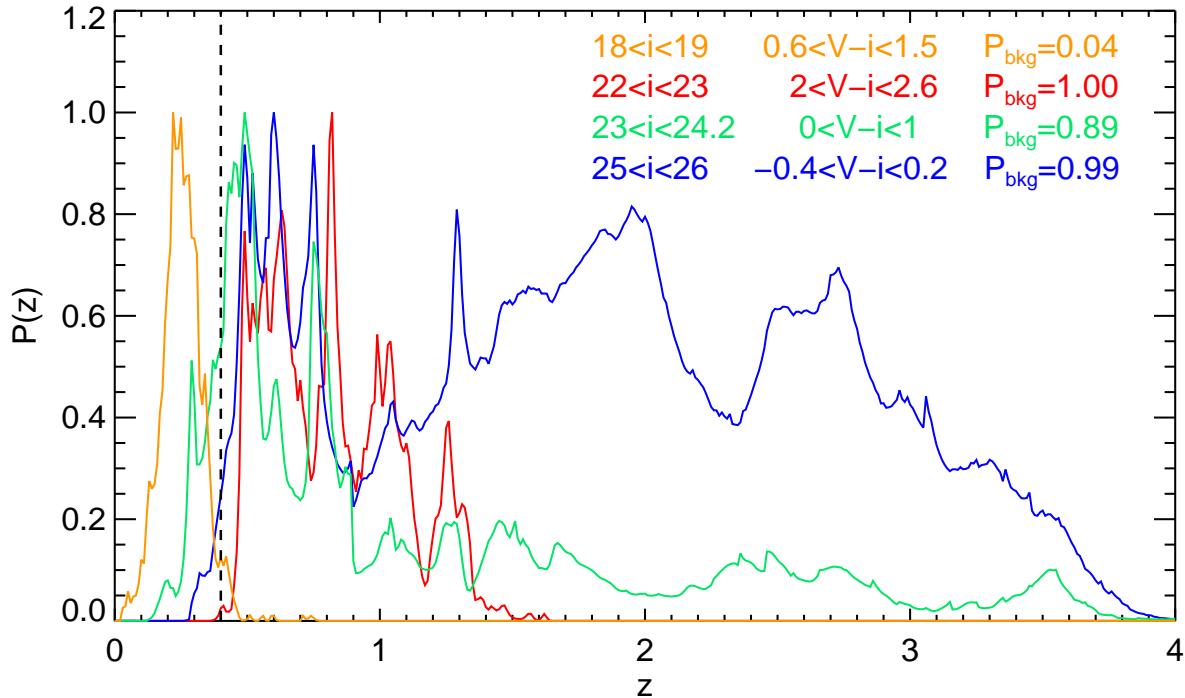}
  }
  \caption{Mean normalised $P(z)$ for different class of
      galaxies, showing in different colours (see legend) red sequence
      galaxies, red galaxies (i.e. above the red sequence), green
    valley galaxies and blue galaxies. The black dashed line, at
    $z=0.4$, shows the threshold at which we select background
    galaxies. $\rm N_{gal}$ indicates the number of galaxies
    populating each $(V-i)/i$ bin, while $\rm P_{bkg}$ is the
    integrated probability for those galaxies to be at $z>0.4$.}
  \label{fig:zphot_comp_col_mag_pdf}
\end{figure*}

We study the photometric redshift distribution of galaxies in the
LoCuSS catalogue as a function of $(V-i)$ colour and $i$-band
magnitude.  First, we split the colour-magnitude plane up into four
regions: red sequence, red, green valley, and faint blue
galaxies.  Red sequence galaxies ($0.6<V-i<1.5$, $18<i<19$) have a
well defined $P(z)$ peak at $z\simeq0.2$ and negligible probability of
being background galaxies, $\pbkgbin=0.04$ (orange curve in
Fig.~\ref{fig:zphot_comp_col_mag_pdf}).  In contrast, red
galaxies ($2.0<V-i<2.6$, $22<i<23$) suffer negligible contamination by
foreground and cluster galaxies, with $\pbkgbin=1.00$ (red curve in
Fig.~\ref{fig:zphot_comp_col_mag_pdf}).  This gives independent
confirmation of \citeauthor{Okabe2013}'s identification of red
galaxies in this colour-magnitude plane as a robust low contamination
strategy for selecting background galaxies.  Note that
\citet{Okabe2013} combined the COSMOS photometric redshift catalogue
with measurements of the reduced shear as a function of $(V-i)$ colour
to define their red colour cut; they did not use the LoCuSS
photometric redshift catalogue discussed here.

Moving blueward of the red sequence, we find that the $P(z)$ of green
valley galaxies ($0<V-i<1$, $23<i<24.2$) peaks at $z\simeq0.5$ and
extends down to $z\simeq0.1$.  Indeed, with $\pbkgbin=0.89$, green
valley galaxies appear to include an appreciable fraction of
foreground and cluster galaxies (green curve in
Fig.~\ref{fig:zphot_comp_col_mag_pdf}).  Finally, faint blue galaxies
($-0.4<V-i<0.2$, $25<i<26$) have a very high probability of being at
$z>0.4$, based on the LoCuSS photometric redshifts, with
$\pbkgbin=0.99$ (green curve in
Fig.~\ref{fig:zphot_comp_col_mag_pdf}).  However the tests discussed
in \S\ref{sec:photoz_COSMOS} indicate that this may be an
over-estimate.

We also bin the colour-magnitude plane more finely to confirm the
stability of these results, finding that the $P(z)$ does not vary
strongly between adjacent bins of width several tenths of a magnitude
(Fig.~\ref{fig:zphot_comp_col_mag_pdf_red} and
Fig.~\ref{fig:zphot_comp_col_mag_pdf_green_blue} of
Appendix~\ref{appendix}).


\subsection{Integrated P(z) as a function of colour and magnitude}

\begin{figure*}
  \centering
  \includegraphics[width=0.68\hsize]{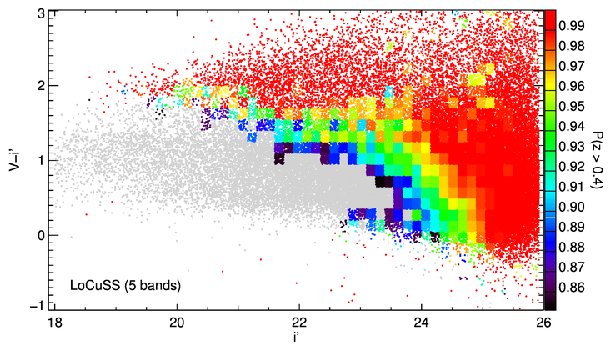}
  \includegraphics[width=0.68\hsize]{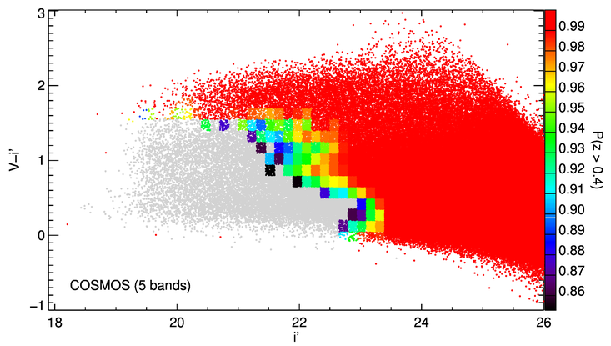}
  \includegraphics[width=0.68\hsize]{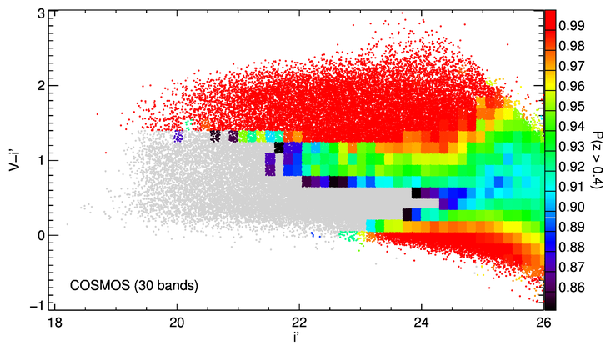}
  \caption{Colour-magnitude diagrams for all sources in the three
    stacked clusters in LoCuSS (top panel) and for COSMOS (middle and
    bottom panels). The colour bar represents the probability that a
    galaxy in a given $(V-i)/i$ bin lies at $z>0.4$ according to the analysis
    based on 5 photometric bands (top and middle panels) and 30 bands
    (bottom panel).  }
  \label{fig:col_mag_p_bkg_bin}
\end{figure*}

We now investigate $\pbkgbin$ as a function of colour and magnitude
for each of the three photometric redshift catalogues
(Fig.~\ref{fig:col_mag_p_bkg_bin}).  We emphasise that the LoCuSS
catalogue benefits from the presence of a massive galaxy
  cluster at $z\simeq0.2$ in the centre of each of three fields of
  view (i.e.\ ABELL\,0068, ABELL\,0383, ABELL\,0611); the COSMOS-30
catalogue lacks massive clusters at $z\simeq0.2$ and benefits from
excellent photometric redshift precision; the COSMOS-5 catalogue
enjoys none of the advantages of the other two catalogues, however it
aides the interpretation of both of the other two catalogues.

The COSMOS-30 catalogue identifies a broad swathe of colour,
$0\ls(V-i)\ls1.5$ as suffering $\gs5\%$ contamination by galaxies at
$z<0.4$, i.e.\ $\pbkgbin\ls0.95$ (bottom panel of
Fig.~\ref{fig:col_mag_p_bkg_bin}).  At bluer colours it
  appears from the COSMOS-30 catalogue that galaxies at $(V-i)\ls0$
  and $i>23$ may suffer contamination at the $\sim1\%$ level --
  i.e.\ competitive with the galaxies at $(V-i)\gs1.5$ that
  \citet{Okabe2013} used.  Turning to the LoCuSS catalogue, we see
that red ($V-i\gs1.5$) and faint ($i\gs24.5$) galaxies suffer
negligible contamination by galaxies at $z<0.4$ (top panel of
Fig.~\ref{fig:col_mag_p_bkg_bin}).  However it is striking that at
$i\gs24$ the LoCuSS $BVRiz$-band photometric redshifts are unable to
pick out the same population of galaxies at $z<0.4$ as seen in the
COSMOS-30 catalogue.  This same feature is also seen in the COSMOS-5
catalogue (top panel of Fig.~\ref{fig:col_mag_p_bkg_bin}), indicating
that it is likely a feature of constraining photometric redshifts with
just five photometric bands.  Nevertheless, a key difference between
the LoCuSS and COSMOS-5 catalogues is that for galaxies bluer than
$(V-i)\ls1.5$ -- i.e. colours expected to suffer noticeable
contamination -- the contamination level only falls below $\sim5\%$
for LoCuSS galaxies at $i\gs24$, in contrast to the same feature
appearing for COSMOS-5 galaxies at $i\gs23$.  We interpret this
difference as being caused by the presence of faint cluster and
foreground galaxies in the LoCuSS catalogue and absence of the same
from the COSMOS catalogues.  Combining the strengths of the
  LoCuSS and COSMOS-30 catalogues therefore suggests that faint blue
  galaxies satisfying $i\gs24$ (based on LoCuSS and COSMOS-30) and
  $(V-i)\ls0.4$ (based on COSMOS-30) should suffer $\ls5\%$
  contamination.


\subsection{Contamination}
\begin{figure}
  \includegraphics[angle=90,width=\hsize]{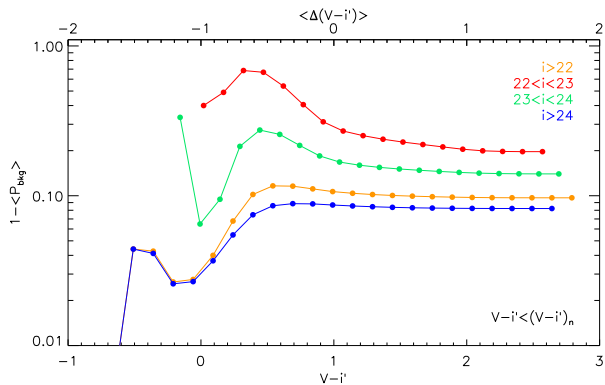}
  \caption{Contamination ($1-\pbkg$, with $\pbkg$ defined in
    Eq.~\ref{p_z04} each galaxy) for all galaxies with $22<i<26$ and
    bluer than a given $(V-i)$ colour, based on the COSMOS-30
    catalogue.  The top axis of each panel shows the typical
    $\Delta(V-i)$ from the red sequence of the three clusters in
    LoCuSS.}
  \label{fig:contamination}
\end{figure}

We now put the preceding discussion on a firmer quantitative
  footing and estimate the fraction of faint blue galaxies that are
  genuine background galaxies.  In the absence of a LoCuSS photometric
  redshift catalogue based on all 30 COSMOS bands, we base our
  estimates on the COSMOS-30 catalogue because it has the most
  accurate photometric redshifts, and rely on the LoCuSS catalogue as
  a sanity check, because it contains clusters at $z\simeq0.2$.
  Essentially the LoCuSS photometric redshifts motivate us to consider
  galaxies fainter than $i=24$.

Based on the COSMOS-30 catalogues we therefore define
contamination as $1-\pbkgbin=1-\langle\pbkg\rangle$ and investigate
how contamination depends on colour and magnitude
(Fig.~\ref{fig:contamination}).  The contamination level for all
galaxies at $i>22$ presents a minimum of $\sim2-3\%$ for the bluest
galaxies, i.e.\ $(V-i)\le0$ (orange curve,
Fig.~\ref{fig:contamination}).  However this is dominated by the
faintest galaxies, i.e.\ $i>24$ (blue curve,
Fig.~\ref{fig:contamination}), with brighter galaxies suffering
contamination upto several tens of per cent (green and red curves,
Fig.~\ref{fig:contamination}).  Our goal of sub-4 per cent
  systematic biases in LoCuSS weak-lensing mass measurements
  (Section~1) therefore appears to be achievable with a galaxy
  selection of $i>24$ and $(V-i)\ls0$.  However this yields $<1$
  galaxy per square arcminute -- i.e.\ a marginal increase on the
  number density achieved by \cite{Okabe2015}.  Moreover given the
  respective limitations of \emph{both} the COSMOS-30 and LoCuSS
  photometric redshift catalogues, the measurement of the systematic
  bias for this blue galaxy selection is less accurate than the
  measurement of 1 per cent contamination of the red galaxies selected
  by \citeauthor{Okabe2015}.  We therefore conclude that including
  blue galaxies in the LoCuSS weak-lensing mass measurements of
  individual clusters is inconsistent with our goal of sub-4 per cent
  control of systematic biases.

To explore the blue galaxy selection issues further, we adopt
  an estimated contamination level of $7$ per cent as a nominal
  threshold for selecting blue background galaxies.  This is
  comparable with the level of uncertainty on measurements of
  systematic biases in other studies in the literature
  \citep[e.g.][]{Applegate2014}.  Whilst this level of contamination
  exceeds our goals within LoCuSS, it allows us to investigate the
  radial distribution of the contaminants.  Based on the analysis
  presented in Figure~5, $7$ per cent contamination by galaxies at
  $i>24$ translates into a colour cut of $V-i<0.4$, corresponding to a
  typical difference in colour from the red sequence
  $\Delta(V-i)=(V-i)-(V-i)_{RS}<-0.6$.  This selection yields a mean
  number density of $3$ galaxies per square arcminute.  We use this
  sample in the following Sections.

\subsection{Reduced shear profile and mass modelling}
\label{sec:wl_analysis}

\begin{figure*}
  \centering \includegraphics[width=0.48\hsize]{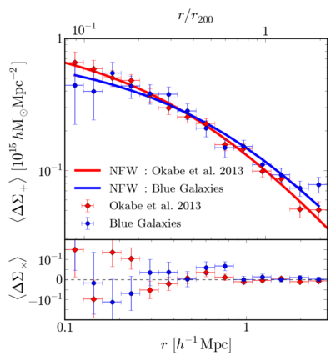}
  \includegraphics[width=0.48\hsize]{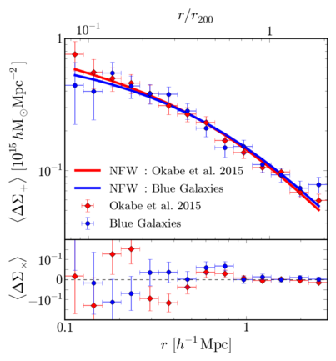}
  \caption{Stacked reduced shear profiles for the full sample of
      50 clusters from LoCuSS, using red galaxy selections as in
      \citet[][left panel]{Okabe2013} and \citet[][right
        panel]{Okabe2015} and the blue selection as described in this
      work (both panels).  The curves represent the best fit NFW halo
      to the respective shear profiles, discussed in
      \S\ref{sec:wl_analysis}.}
  \label{fig:Ok_fig3}
\end{figure*}

We now probe the mass distribution of the High-$L_X$ sample using the
blue background galaxy sample, defined by $i>24$, $\Delta(V-i)<-0.6$,
and using the galaxy shape measurements from \cite{Okabe2015}.  These
shape measurements are based on methods introduced by \cite[the
  so-called KSB method]{KSB1995} using a modified version of the {\sc
  imcat} package \citep{Okabe2013, Okabe2015}.  Key features include
that we test the reliability of the shape measurement pipeline using
simulations that match our observational data and targets.  In
particular we test our code in the high-shear regime, $g\simeq0.3$,
and simulate data with the same field of view as Suprime-CAM such that
we can include sufficient galaxies detected at high signal-to-noise
ratio ($\nu>30$) to test our approach to measuring the isotropic point
spread function correction.  This is essential to confirm our ability
to measure reliable galaxy shapes for galaxies with $i\simeq25$.
Indeed, the multiplicative bias in our shape measurements is just a
few per cent and independent of apparent magnitude.  Full details are
provided in \citet{Okabe2013} and \citet{Okabe2015}; the latter uses a
modified version of the former's shape measurement pipeline. In
  this article we use the same COSMOS UltraVISTA photometric redshift
  catalogue as \citeauthor{Okabe2015}, and also use their faint galaxy
  shape measurements.  

We compute the stacked ``blue galaxy'' reduced shear profile for the
full sample of 50 clusters, centred on the respective brightest
cluster galaxies (BCGs) following the same approach as
\citet{Okabe2013} (Fig.~\ref{fig:Ok_fig3}).  Our blue galaxy shear
profile agrees well with \citeauthor{Okabe2013}'s red galaxy shear
profile at $0.2-2h^{-1}\Mpc$, with a slightly lower blue shear signal
on smaller scales, and larger signal on larger scales.  To quantify
the level of agreement, we fit a Navarro, Frenk, \& White
(\citeyear[][NFW]{NFW1997}) profile to the blue shear profile,
obtaining $M_{200}=(7.68_{-0.67}^{+0.71})\times10^{14} h^{-1}
M_\odot$, $c_{200}=3.06_{-0.28}^{+0.30}$, i.e.\ higher mass and lower
concentration than \citet{Okabe2013} obtained with red galaxies at
$\sim2\sigma$ significance.  However a more faithful
like-for-like comparison is between our blue galaxy fit and a fit
  to the new \cite{Okabe2015} red background galaxy sample, because it
  is based on the same updated version of the shape measurement
  pipeline and the same COSMOS UltraVISTA photometric redshift
  catalogue.  Note that our fit to the blue galaxies and the
  \citet{Okabe2013} fit discussed above both fit a single NFW halo to
  the data.  \citet{Okabe2015} fit a model comprising an NFW halo, a
  point mass to represent the brightest cluster galaxy, and a two-halo
  term to their red galaxy shear profile.  Here, to facilitate a
  direct comparison with \citet{Okabe2013} and our blue galaxy shear
  profile fit, we fit solely an NFW halo to \citeauthor{Okabe2015}'s
  red galaxy shear profile.  We obtain $M_{200}= (7.14^{+0.44}
_{-0.42})\times10^{14}h^{-1}M_\odot$, $c_{200}=3.39^{+0.26} _{-0.25}$,
in excellent agreement with our blue galaxy result -- the discrepancy
is reduced to $\ls1\sigma$. The better agreement in the NFW model
  fits can be traced to the good agreement of the blue and red shear
  profiles on scales of a few Mpc (Fig.~\ref{fig:Ok_fig3}).  

\subsection{Faint blue galaxy distribution}

\begin{figure*}
\centering
   \includegraphics[trim=0cm 1cm 0cm 2cm,clip=true,width=0.48\hsize]{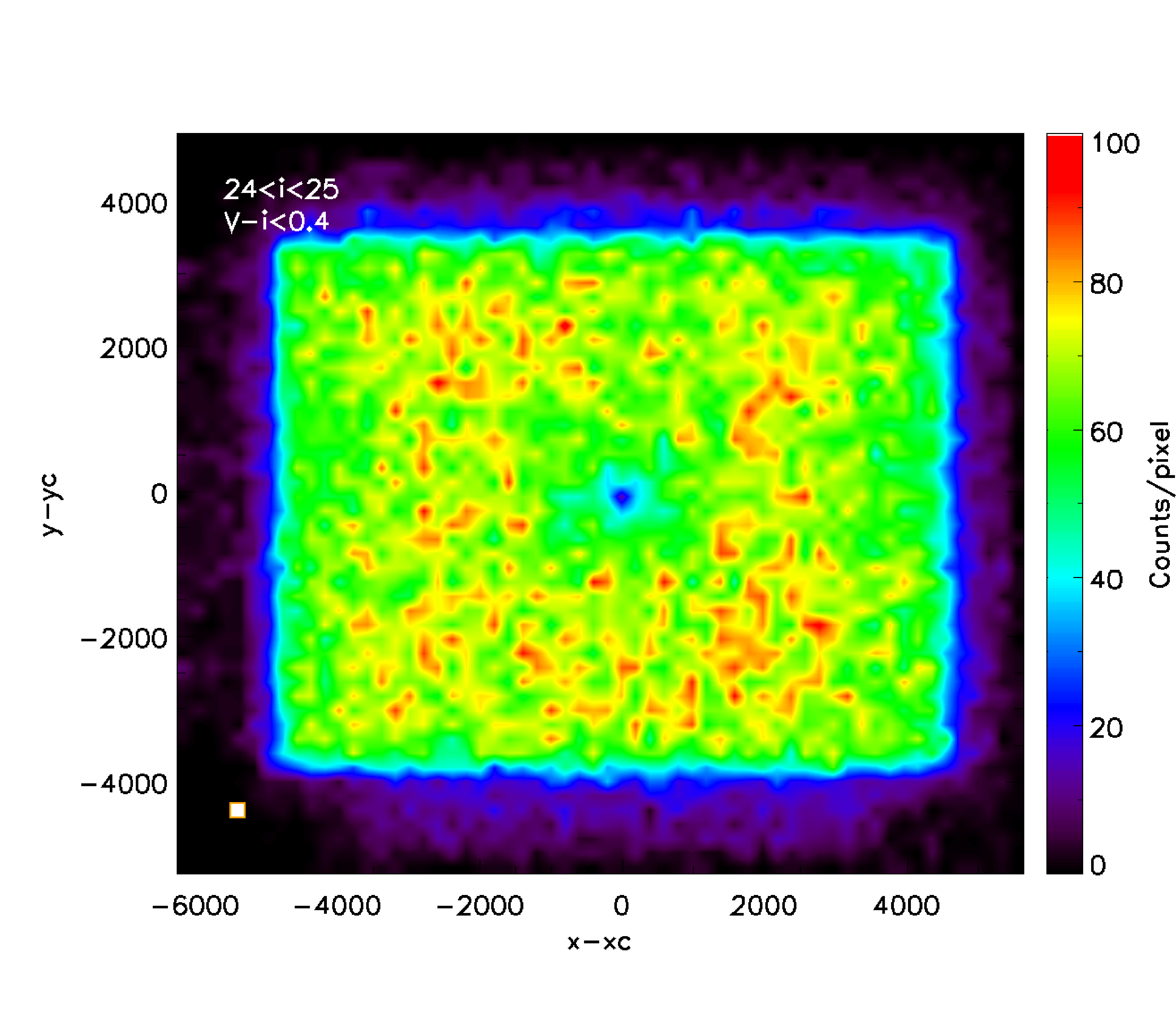}      
   \includegraphics[trim=0cm 1cm 0cm 2cm,clip=true,width=0.48\hsize]{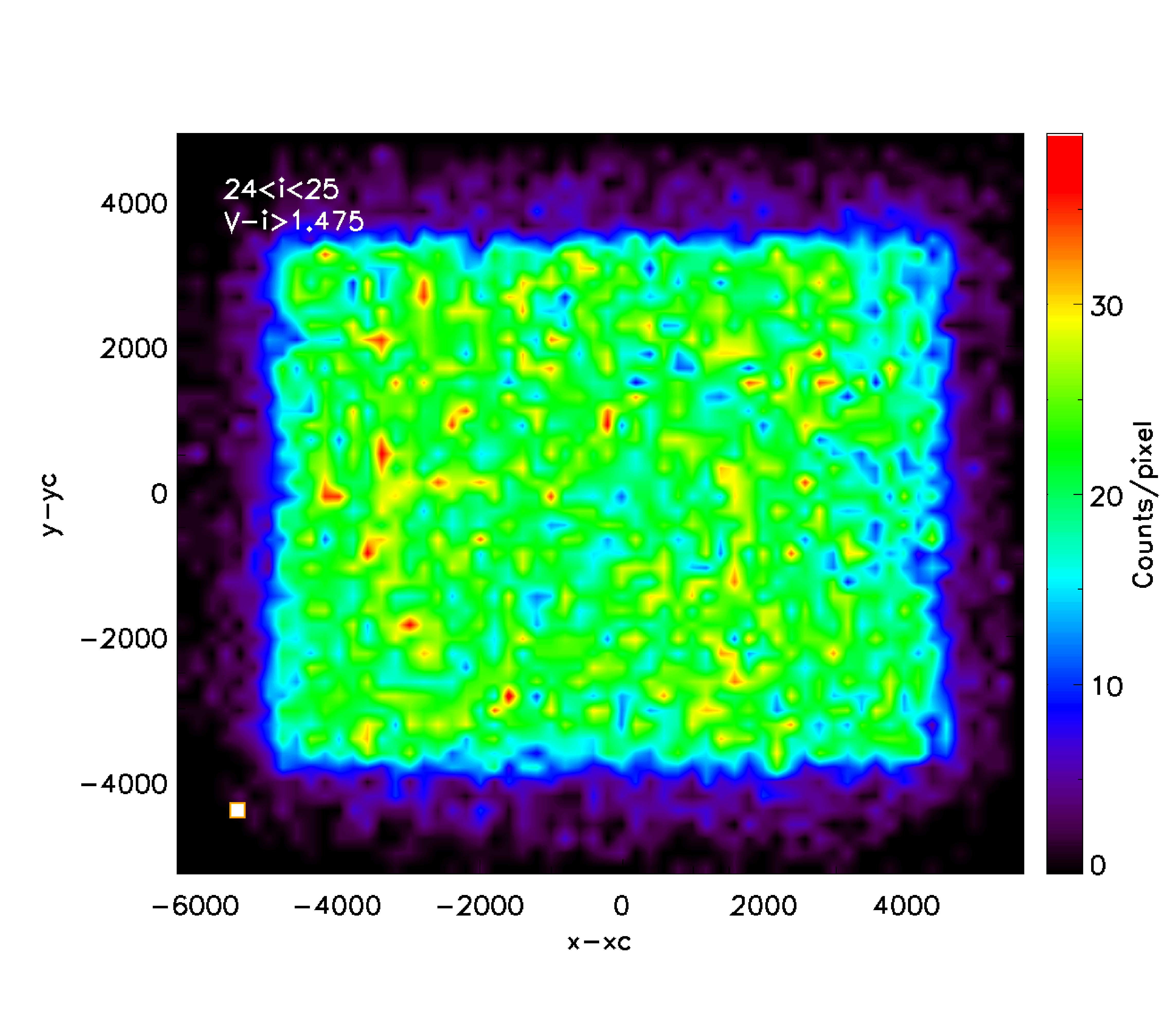}      
  \caption{Density map of faint ($24<i<25$) blue ($V-i<0.4$, left
    panel) and red ($V-i>1.475$, right panel) galaxies with a shape
    measurement for 50 clusters. The orange and white square at the
    bottom left of each panel shows the bin size used
    to compute the density maps. The colour bar indicates the number
    of galaxies per pixel in each panel. }
  \label{fig:faint_blue_distr}
\end{figure*}

\begin{figure}
  \centerline{\hspace{-7mm}
  \includegraphics[angle=-90,width=1\hsize]{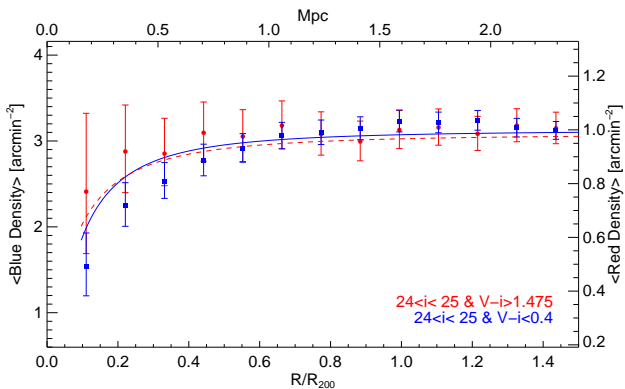}}
  \caption{Stacked number density profile of red and blue galaxies
    behind 50 clusters in the High-$L_X$ cluster sample.  Red-dashed
    line and blue solid line represent the prediction for red and blue
    galaxies, respectively, based on the magnification modelling
    described in the text. The two profiles are normalised to the mean
    density of blue galaxies at $\rm R>R_{200}$.}
  \label{fig:dens_profile}
\end{figure}

We further test the reliability of the blue background galaxy
catalogue by examining the angular distribution of these galaxies.
Ignoring the effects of lensing magnification, the distribution of
background galaxies should display no correlation with the location of
the foreground galaxy clusters.  We find that the observed number
density of blue galaxies drops towards the centre of the clusters, and
any drop in the number density of red galaxies (defined as in
\citealt{Okabe2013}) is much less pronounced
(Fig.~\ref{fig:faint_blue_distr}).  A more quantitative view is
obtained from the stacked cluster centric number density profile of
the blue and red galaxy samples, revealing that the radial
distribution of red galaxies is consistent with being flat and a
statistically significant drop is detected in the blue galaxy profile
(Fig.~\ref{fig:dens_profile}).  To interpret this behaviour robustly
requires consideration of the lensing magnification, which acts to
boost the sensitivity of the observations through the central cluster
regions, at the expense of reducing the intrinsic solid angle that is
probed by the data.  We compute the run of number density of
background galaxies expected from an intrinsically flat distribution
of galaxies located behind the clusters, adopting the best-fit NFW
profile to the red galaxies as our best estimate of the mean cluster
mass distribution.  We compute the density ($\kappa$) and shear
($\gamma$) profiles of this model using the formalism provided by
\cite{WrightBrainerd2000} and from these profiles obtain the
magnification profile:
\begin{equation}
  \mu(\theta)=\frac{1}{(1-\kappa(\theta))^2-\gamma(\theta)^2}.
\end{equation}
The magnification profile encodes how the presence of the foreground
clusters modify both the intrinsic solid angle probed by the
observations and the photometric depth of the data.  The former is a
straightforward factor by which we multiply the solid angle of each
radial bin.  The latter is taken account of by adjusting the
photometric limit down to which we count galaxies as a function of
cluster centric radius:
\begin{equation}
  i_{\rm limit}(\theta)=i_0+2.5\log(\mu(\theta))
\end{equation}
where $i_0$ is the unlensed depth of the data, and $i_{\rm limit}$ is
the limit after correcting for lensing magnification.  To ensure the
accuracy of our calculations we implement two additional steps in the
selection of background galaxies.  Firstly, we
select galaxies from those for which \cite{Okabe2015} have measured a
shape, as opposed to the full $V/i$-band photometric catalogue.
Secondly, we restrict the selection of observed galaxies in the regime
of negligible magnification (i.e.\ $\mu=1$) to $i_0=25$, i.e. a full
magnitude brighter than the nominal depth of our data.  This is
important, to ensure that when we compute the expected number density
of galaxies at $\mu>1$, we do so down to a value of $i_{\rm limit}$
that does not suffer incompleteness. 
Note that at projected cluster
centric radii of $R\gs150\kpc$ we obtain $i_{\rm limit}-i_0\ls0.5$
magnitudes.  

The magnification bias is expressed in terms of the number density of
background galaxies as $n_\mu = n_0 \mu ^{2.5s -1}$ where $n_0$ is the
unlensed mean surface number density of background galaxies, $\mu$ is
the magnification and $s$ is the slope of the number counts of
galaxies in the cluster outskirts ($R>R_{200}$). For the faint limit
used in this work ($i\sim25$) we find a slope $s=0.12$ for red
galaxies and $s=0.22$ for blue galaxies. In both cases, the slopes are
less than the lensing invariant slope $s=0.4$ implying a deficit of
background galaxies. 

Both red and blue galaxies are consistent with the model prediction
within the errors (Fig.~\ref{fig:dens_profile}), albeit with a
possible deficit of blue galaxies relative to the model at
$R\ls0.5\Mpc$.  However the amplitude of this possible deficit is
comparable with the fractional decrease in the observed solid angle
per bin due to obscuration by bright cluster galaxies
\citep{Umetsu2014}, we therefore do not regard this difference as
significant.  Note that we have not corrected the data points in
Fig.~\ref{fig:dens_profile} for this effect.  

The apparently flat profile of red galaxies in
Fig.~\ref{fig:dens_profile} might suggest a tension with the
typical use of red galaxies for count-depletion measurements,
\cite[e.g.][]{Medezinski2013,Umetsu2012}.  In fact, these authors use
a population of faint ($z'\lesssim 25$), $R-z'$ red ($R-z'\sim1$), but
$B-R$ blue ($B-R\lesssim 0.5$) galaxies from deep $BRz'$ imaging
(e.g., \citealt{Medezinski2013}, figure 2, table 3;
\citealt{Umetsu2012}, figure 3, table 3).  Although the $N(z)$ of
their $BRz'$ red galaxies peaks around $z=1$ \cite[their
  figure7]{Umetsu2010} there might be an overlap with our blue
population (see Fig.~\ref{fig:zphot_comp_col_mag_pdf}), i.e.\ the red
galaxies selected on $(V-i)/i$ colours are not necessarily the same
population of galaxies selected in the $BRz'$ space.

Finally, we note that the magnification model curves for both red and
blue galaxies show a gentle dependence on cluster centric radius.
Based on our data and sample, we find that the observed number density
of galaxies that obey an intrinsically flat radial distribution
declines by a factor of $1.29$ between $2\Mpc$ and $0.5\Mpc$ and a
factor of 1.07 between $3\Mpc$ and $0.75\Mpc$ -- i.e.\ on the
respective scales on which \citet{Hoekstra2012} and
\citet{Applegate2014} both assume that the \emph{observed} number
  density profile of background galaxies is flat.


\section{Summary and Discussion}
\label{sec:discussion}

We have tested the reliability of photometric redshifts as a tool for
selecting background galaxies for galaxy cluster weak-lensing mass
measurements.  Our main motivation is that contamination of background
galaxy samples by faint foreground and cluster galaxies is one of the
major sources of systematic uncertainty in this field.  In particular,
we investigate the robustness of photometric redshifts based on five
photometric filters (that is common on the literature) for cluster
weak-lensing, the dependence of our results on galaxy colour, and run
of background galaxy number density with clustercentric radius.  These
tests are important for our own programme of weak-lensing mass
measurements using $V$- and $i$-band data within the Local Cluster
Substructure Survey (LoCuSS), and generally applicable to other
surveys that select from colour-colour planes, one or more
colour-magnitude planes, and/or photometric redshift catalogues based
on five photometric bands
\citep[e.g.][]{High2012,Applegate2014,Hoekstra2015}.  
In summary, our main results are as follows:
\begin{enumerate}
\item We use a deep spectroscopic catalogue from ACReS and the COSMOS
  catalogue to test our photometric redshifts that are derived from
  $BVRiz$-band Subaru observations of three galaxy clusters at
  $z\simeq0.2$.  Our redshifts are accurate in the mean to
  $\sigma\simeq0.04$, suffer negligible bias $|\mu_{\Delta
    z}|\ls0.01$, and a catastrophic failure rate of $\eta\ls0.1$.
  This is competitive with other surveys that derive photometric
  redshifts from similar data for the purpose of selecting background
  galaxies.
\item We confirm that the selection of galaxies redder than the red
  sequence of cluster members by $\Delta(V-i)\ge0.475$ suffer sub per
  cent contamination by faint foreground and cluster galaxies, in
  particular galaxies at $z_{\rm phot}<0.4$.  This provides
  independent support for the red galaxy selection methods developed
  by \citet{Okabe2013}.
\item In contrast, faint blue galaxies are difficult to place
  accurately along the line of sight through clusters due to our
  inability to break the degeneracy between the spectral shape of blue
  galaxies at $z\ls0.3$ and at $z\simeq2-3$.  This highlights the
  importance of deep $u$-band photometry.
\item We compare LoCuSS photometric redshifts and COSMOS photometric
  redshifts and find that neither are adequate on their own to
  identify a low contamination region of $(V-i)/i$ colour magnitude
  space for selection of blue background galaxies.  The LoCuSS
  redshifts suffer from limited photometric bands, and whilst the
  COSMOS redshifts benefit from 30 photometric bands, the COSMOS field
  does not contain any massive galaxy clusters at $z\simeq0.2$.  This
  is an important result, because it is common in the literature
  either to use the COSMOS catalogue to calibrate the reliabilty and
  redshift distribution of colour-selected galaxy
  samples \citep[e.g.][]{Okabe2010}, or to base the selection of
  background galaxies purely on photometric redshifts derived from 5
  photometric filters \citep[e.g.][]{Applegate2014}.  Similarly,
  photometric redshift catalogues based on 5 filter observations of
  blank fields, for example the 5-band COSMOS catalogue that we
  compute, or the CFHTLenS catalogue used by \citet{High2012} and
  \citet{Hoekstra2012} suffer from limited photometric coverage and
  absence of clusters from the photometric redshift catalogue.
\item We combine the strengths of the LoCuSS and COSMOS photometric
  redshift catalogues to identify a region of colour-magnitude space
  at $i>24$, $(V-i)<0.4$ (corresponding to $0.6$ magnitudes bluer than
  the typical cluster red sequence at $z\simeq0.2$) in which data and
  spectral models are consistent with contamination of colour-selected
  blue background galaxy samples being $\sim7$ per cent.  We
    apply this blue galaxy selection to measure the stacked reduced
    shear profile of the full High-$L_X$ LoCuSS cluster sample.  The
    best-fit \cite{NFW1997} density profile has a mass and
    concentration that is consistent with the red galaxy shear
    profile-based results of \cite{Okabe2013} within $\sim2\sigma$ and
    \cite{Okabe2015} within $\ls1\sigma$.  The comparison between our
    results and the latter study is more like-for-like than with the
    former.
\item We further explore contamination by examining the radial number
  density profile of colour-selected blue and red galaxies, obtaining
  a red galaxy profile that is consistent with being flat across the
  full range of clustercentric radii probed, and a blue galaxy profile
  that dips by a factor of $\sim2$ towards the central cluster region.
  Both number density profiles are consistent with that expected from
  consideration of the lensing magnification of the foreground
  clusters and the slope of the number counts of faint red and blue
  galaxies measured securely brighter than our detection limit.  Our
  data are therefore consistent with negligible radial trend in the
  level of contamination suffered by red and blue galaxy samples.
\item Recent studies have corrected the slope of their observed number
  density profile of background galaxies to be radially flat whilst
  implicitly assuming that lensing magnification has a
  negligible affect on the number density profile.  Our analysis
  suggests that this assumption is not valid.  Certainly,
  for our sample and data, we find that the observed number density of
  galaxies that obey an intrinsically flat radial distribution
  declines by a factor of $1.29$ between $2\Mpc$ and $0.5\Mpc$ and a
  factor of $1.07$ between $3\Mpc$ and $0.75\Mpc$ -- i.e. the radial
  ranges adopted by \citet{Hoekstra2015} and \citet{Applegate2014}
  respectively.
\end{enumerate}

In the context of LoCuSS, our specific goal was to
  investigate the feasibility of including faint blue galaxies in the
  LoCuSS weak-lensing analysis whilst maintaining the systematic bias
  from contamination at a level sub-dominant to the statistical
  uncertainties.  With a sample of 50 clusters, and typical
  weak-lensing mass measurement error of 30 per cent, this equates to
  aiming to control contamination at the sub-4 per cent level.  Given
  the absence of massive galaxy clusters from the COSMOS-30
  photometric redshift catalogue, and the shortcomings of 5-band
  photometric redshift catalogues of clusters that we (and other
  surveys) have at our disposal, we concluded that this goal is not
  achievable.  \cite{Okabe2015} therefore base their weak-lensing mass
  measurements solely on red galaxies, taking advantage of their new
  radially-dependent red galaxy selection to achieve 13 galaxies per
  square arcminute and 1 per cent contamination.

\section*{Acknowledgments}

We thank Olivier Ilbert and Peter Capak for making the latest version
of the photometric COSMOS catalogue available.  We thank Keiichi
Umetsu, Sean McGee, Walter del Pozzo and Trevor Sidery for useful
discussions, and thank Maggie Lieu and Rossella Martino for
assistance.  We also acknowledge stimulating and cordial
  discussions with Douglas Applegate, Anja von der Linden, Adam Mantz,
  and Henk Hoekstra.  FZ and GPS acknowledge support from the Science
and Technology Facilities Council.  GPS acknowledges support from the
Royal Society. NO (26800097) is supported by a Grant-in-Aid from the
Ministry of Education, Culture, Sports, Science, and Technology of
Japan, and Core Research for Energetic Universe in Hiroshima
University (the MEXT program for promoting the enhancement of research
universities, Japan).  This work was supported by \textquotedblleft
World Premier International Research Center Initiative (WPI
Initiative)\textquotedblright \, and the Funds for the Development of
Human Resources in Science and Technology under MEXT, Japan.  CPH was
funded by CONICYT Anillo project ACT-1122.

\appendix
\section{P(z) as a function of colour and magnitude}

\label{appendix}

\begin{figure*}
\centering
   \includegraphics[angle=-90,width=0.95\hsize]{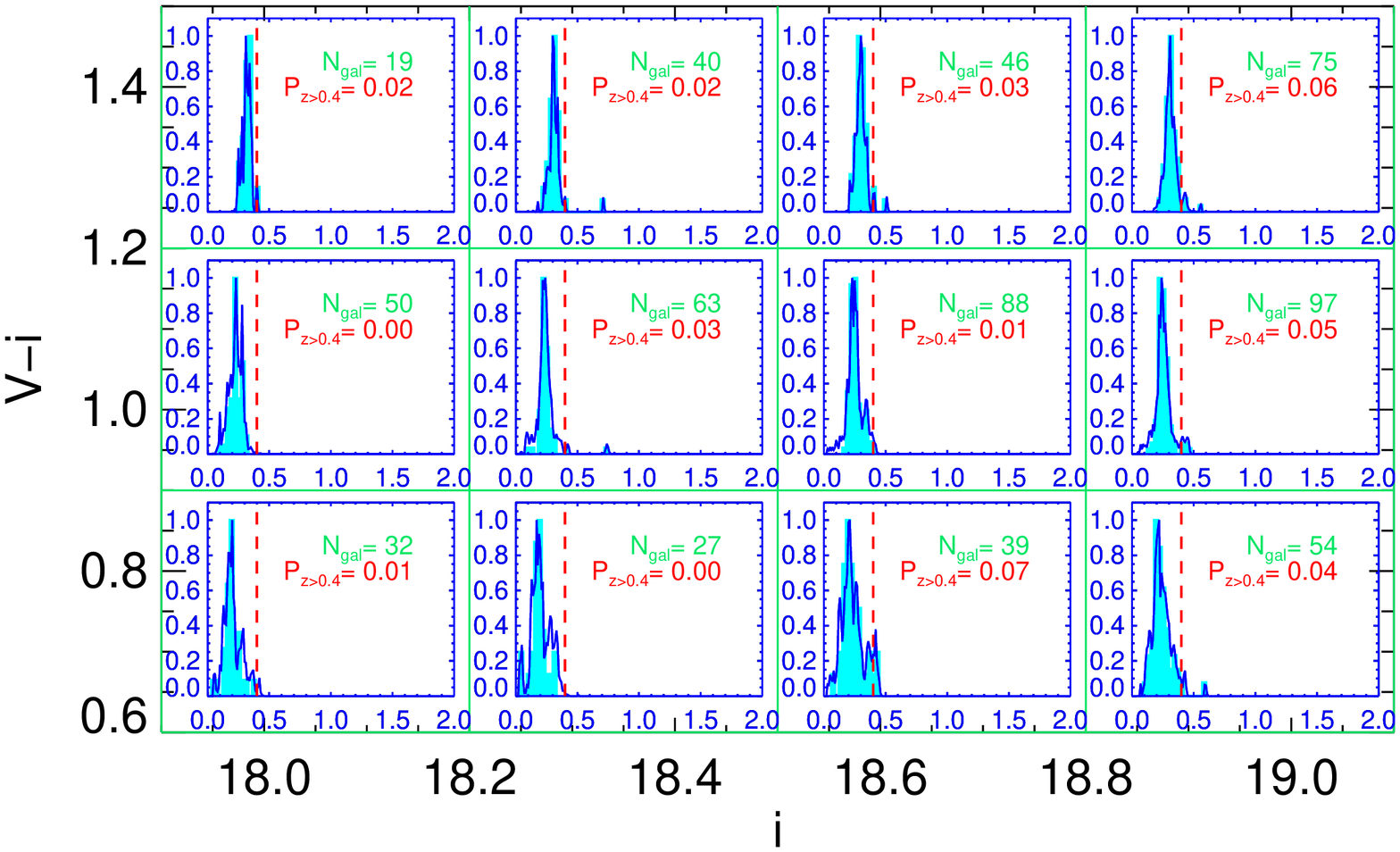}
   \includegraphics[angle=-90,width=0.95\hsize]{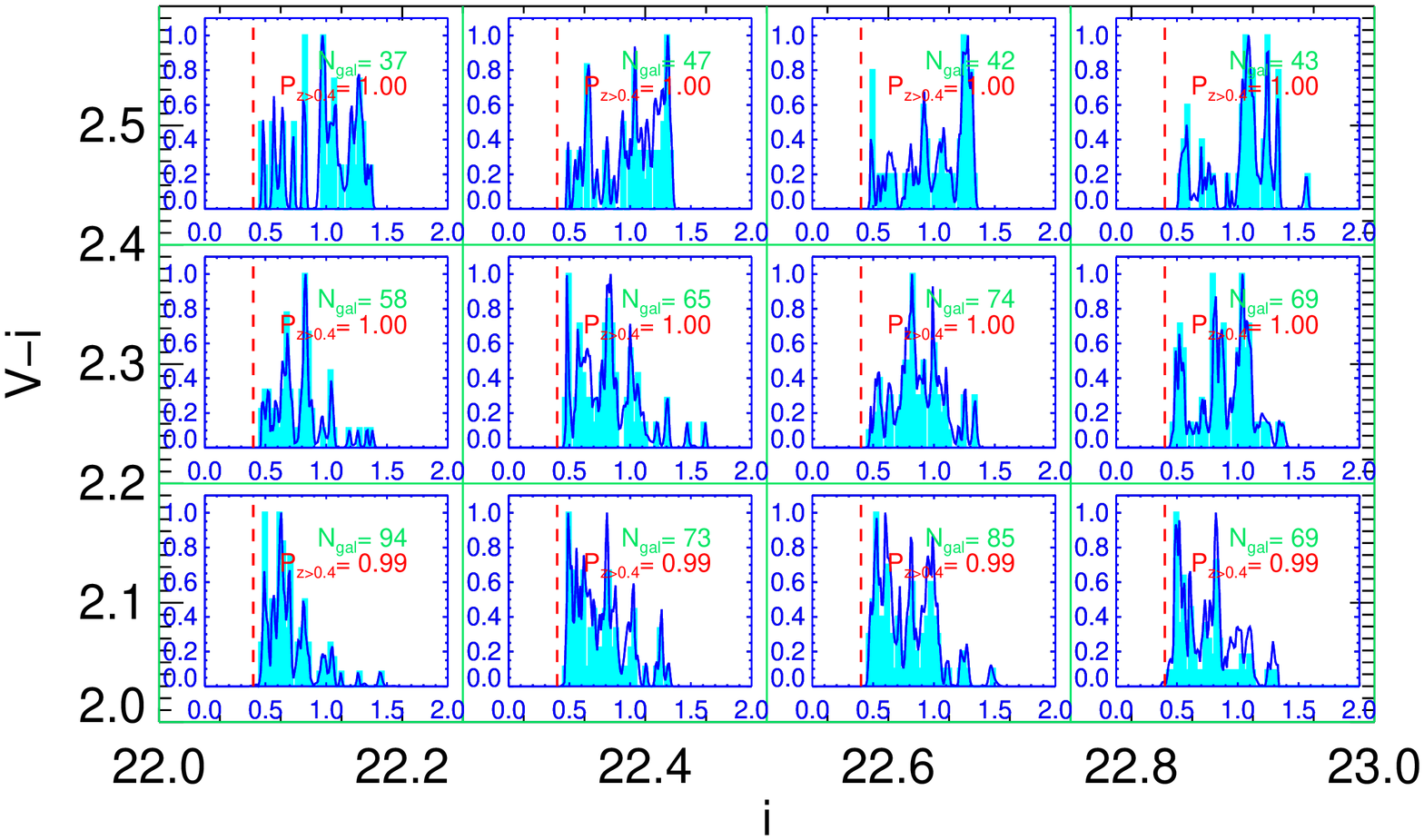}
   \caption{Multi-dimension analysis in the colour-magnitude space for red sequence (top panel) and red galaxies (i.e. above the red sequence, bottom panel). The blue solid line indicates the average P(z) as a function of redshift in the colour-magnitude cell, while the cyan histogram represents the normalised redshifts distribution of all galaxies in a given cell. The red dashed line, at $ z=0.4$, shows the threshold at which we select background galaxies. $\rm N_{gal}$ indicates the number of galaxies populating each cell, while $\rm P_{z>0.4}$ is the integrated probability for those galaxies to be at $z>0.4$.} 
  \label{fig:zphot_comp_col_mag_pdf_red}
\end{figure*}

\begin{figure*}
\centering
   \includegraphics[angle=-90,width=0.95\hsize]{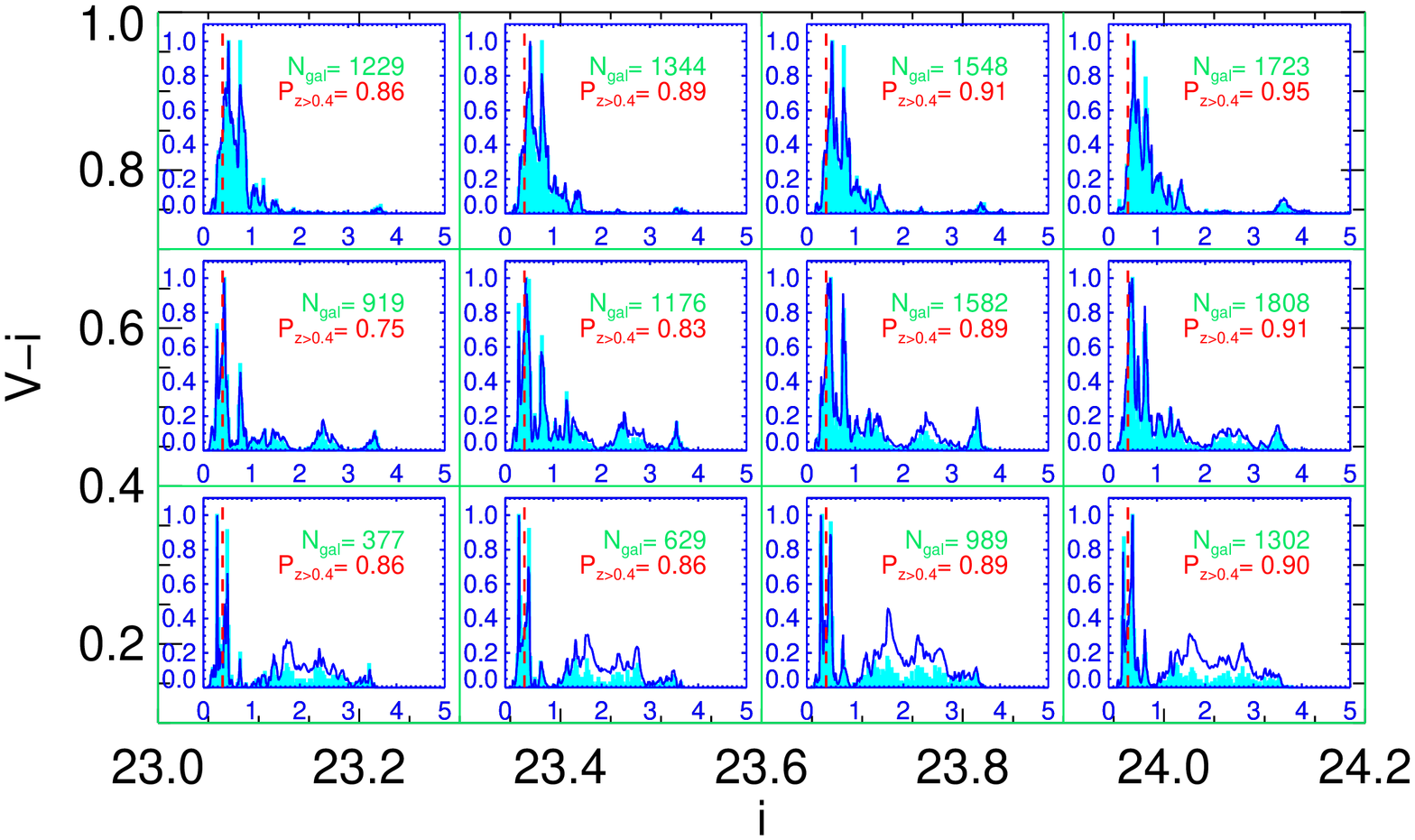}
   \includegraphics[angle=-90,width=0.95\hsize]{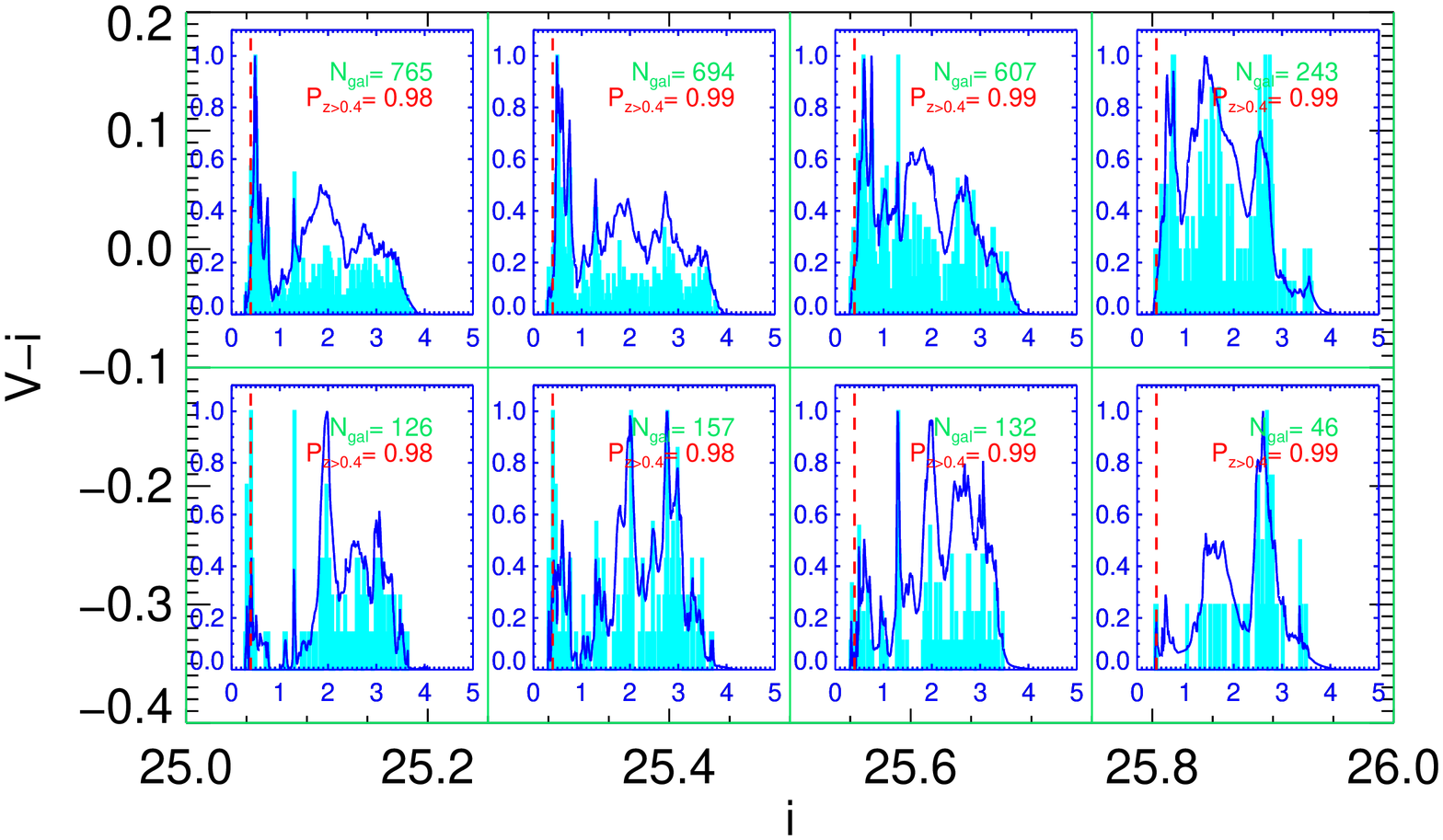}
   \caption{Same as Fig.~\ref{fig:zphot_comp_col_mag_pdf_red} but for green valley galaxies (top panel) and faint blue galaxies (bottom panel). 
   } 
  \label{fig:zphot_comp_col_mag_pdf_green_blue}
\end{figure*}

\bsp

\label{lastpage}

\end{document}